\documentclass[11pt,a4paper]{article} 
\usepackage{amsmath,amssymb,amsfonts}
\usepackage[normal,font=small,labelfont=bf,labelsep=period]{caption}
\usepackage[pdftex]{color,graphicx}
\usepackage[english]{babel}
\usepackage[compress]{cite}
\usepackage{afterpage}
\usepackage[multiple]{footmisc}
\usepackage[dvipsnames]{xcolor}
\definecolor{darkgreen}{rgb}{0,0.65,0}
\usepackage{slashed}
\usepackage[linktoc=all,colorlinks=true,linkcolor=black,urlcolor=magenta,citecolor=black,hyperfootnotes=false]{hyperref}
\hypersetup{%
  colorlinks = true,
  linkcolor  = black
}
\begin{document}
\begin{flushright}
\footnotesize
{IFT-UAM/CSIC-19-39, MPP-2019-75, DESY 19-056, TUM-HEP-1196-19}
\end{flushright}
\vspace{1cm}
\begin{center}
{\LARGE\color{black}\bf Several Problems in Particle Physics and Cosmology Solved in One SMASH \\[1mm] }
\medskip
\bigskip\color{black}\vspace{0.6cm}
{
{\bf Guillermo Ballesteros,$^{1,2}$ Javier Redondo,$^{3,4}$ Andreas Ringwald$^{5,*}$ and Carlos Tamarit$^{6}$}
}
\\[7mm]
\small{$^{1}$Instituto de F\'isica Te\'orica, UAM/CSIC, Cantoblanco, Madrid, Spain  \\
{$^2$Departamento de F\'isica Te\'orica, Universidad Aut\'onoma de Madrid (UAM). Cantoblanco, Madrid, Spain}\\
$^{3}$Departamento de Fisica Teorica, Universidad de Zaragoza, Zaragoza, Spain \\
$^{4}$Max-Planck-Institut f\"ur Physik, M\"unchen, Germany \\
$^{5}$Deutsches Elektronen-Synchrotron DESY, Hamburg, Germany \\
$^{6}$Physik Department T70, Technische Universit\"at M\"unchen, 
Garching, Germany}
\end{center}
\vspace{1cm}
\begin{abstract}
The Standard Model (SM) of particle physics is a big success. However, it lacks 
explanations for cosmic inflation,  the matter-anti-matter 
asymmetry of the Universe,  dark matter,  neutrino oscillations, and  the feebleness 
of CP violation in the strong interactions. The latter may be explained by a complex scalar field charged under a spontaneously broken global U(1) Peccei-Quinn (PQ) symmetry.
Moreover, the pseudo Nambu-Goldstone boson of this breaking -the axion- 
may play the role of the dark matter. Furthermore, the modulus of the PQ field is a candidate for driving inflation. If additionally three extra SM  singlet neutrinos (whose mass is induced by the PQ field) are included, the five aforementioned problems can be addressed at once. We review the SM extension dubbed SMASH --for SM-Axion-Seesaw-Higgs portal inflation--,  discuss its predictions  
and tests in astrophysics, cosmology, and laboratory experiments. Variants of SMASH are also considered and commented on. 
\end{abstract}
\begin{center} 
\vfill\flushleft
\noindent\rule{6cm}{0.4pt}\\
{\small  \tt $^*$ andreas.ringwald@desy.de}
\end{center}

\newpage

\section{INTRODUCTION}

The SM describes the interactions of all known elementary particles with remarkable accuracy. 
Collider and other particle physics experiments have seen so far no significant deviation from its predictions. 
However, there are  fundamental problems in particle physics and cosmology that require the existence of new physics beyond the SM.  Most importantly, there is highly compelling evidence, ranging from the shapes of the rotation curves of spiral galaxies to the temperature fluctuations of the cosmic microwave background (CMB), that almost 85 \% 
of the matter in the Universe is non-baryonic. Moreover, the SM cannot generate the primordial exponential 
expansion of the Universe called inflation that is needed to explain the statistically isotropic, Gaussian and nearly scale invariant temperature fluctuations of the CMB. The SM also lacks enough CP violation to explain why the Universe appears to contain a much larger fraction of baryonic matter than of anti-matter. Furthermore, in the SM, neutrinos are massless, but (tiny) masses are required for the explanation of the observed neutrino flavour oscillations. Last, but not least, the SM suffers from the strong CP problem: it does not explain the smallness of the $\overline\theta$-angle of quantum chromodynamics (QCD) which induces CP-violation in flavour-diagonal interactions, notably a non-zero electric dipole moment of the neutron. In fact, the non-observation of the latter leads to the very strong upper limit $|\overline\theta| < 10^{-10}$,  requiring an extreme fine-tuning which cannot even be justified on the basis of anthropic arguments.

\begin{figure}[t]
\begin{center}
\includegraphics[width=0.6\textwidth]{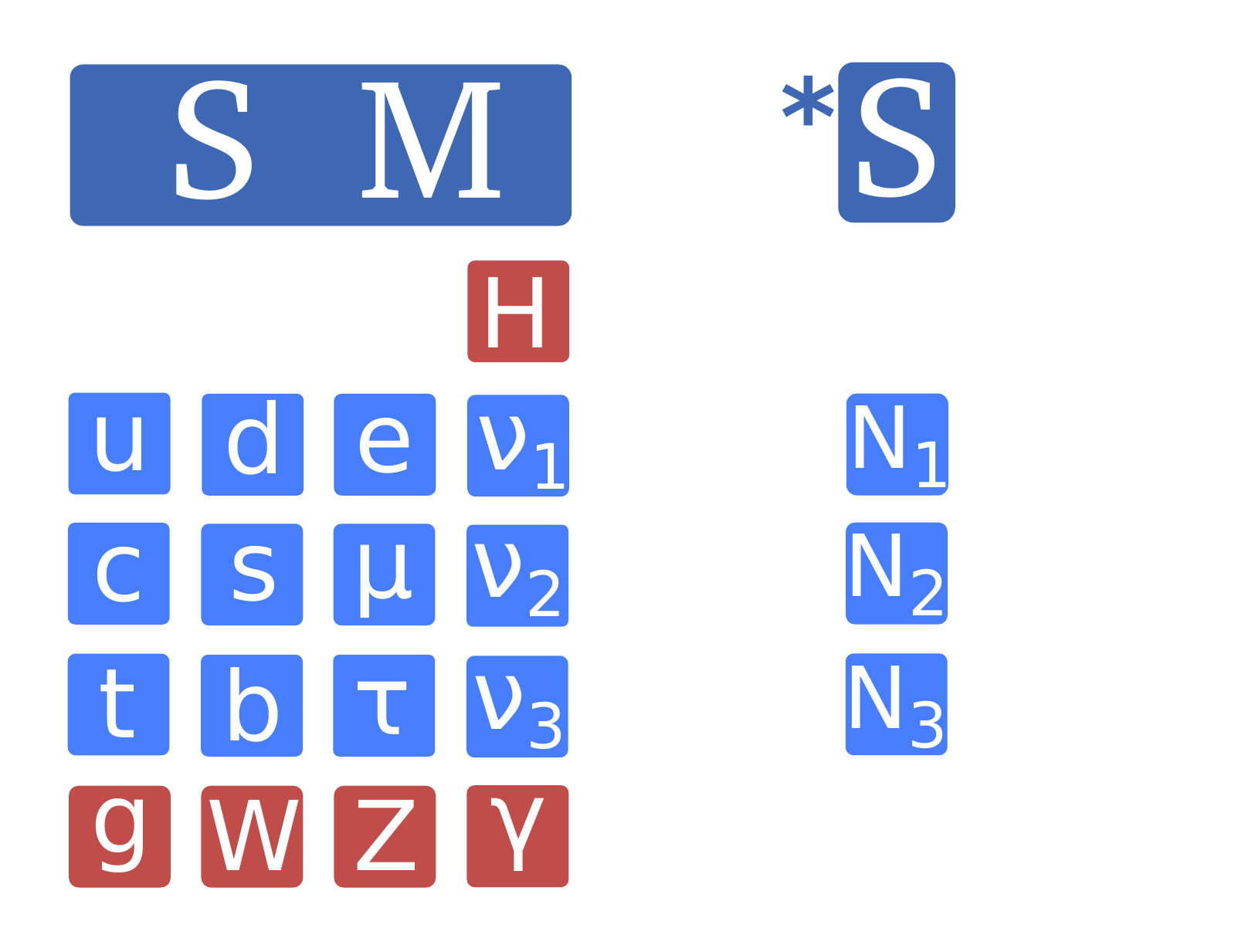}
\caption{\label{fig:nuMSM} Particle/field content of the $\nu$MSM. 
}
\end{center}
\end{figure}

Three of these problems  can be tackled simultaneously in 
the Neutrino Minimal SM ($\nu$MSM)~\cite{Asaka:2005an,Asaka:2005pn}: a remarkably simple extension of the SM by three right-handed singlet neutrinos
$N_i$ (cf. Fig. \ref{fig:nuMSM}),
having Dirac masses $m_D=F v/\sqrt{2}$ arising from Yukawa couplings $F$ with the Higgs ($H$) and lepton ($L_i$)
doublets, as well as explicit Majorana masses $M$, 
\begin{equation}
  {\mathcal L}\supset -\big[ F_{ij}L_i\epsilon H N_j+\frac{1}{2}M_{ij} N_i  N_j \big]\,,
\label{yuk_nuMSM}
\end{equation}
(in Weyl spinor notation).  
In the seesaw limit, $M\gg m_D$, the neutrino mass spectrum splits into a light set given by the eigenvalues 
$m_1<m_2<m_3$ of 
the matrix 
\begin{equation}
\label{eq:seesaw_neutrino_masses}
m_\nu = - m_D M^{-1} m_D^T\,,
\end{equation} 
with the eigenstates corresponding mainly to mixings of the active 
left-handed neutrinos $\nu_\alpha$, and a heavy set given by the eigenvalues $M_1<M_2<M_3$ of the matrix $M$, with the eigenstates corresponding to mixings of the sterile right-handed neutrinos $N_i$. The neutrino mass and mixing problem is thus solved by the usual type-I seesaw  mechanism~\cite{Minkowski:1977sc,GellMann:1980vs,Yanagida:1979as,Mohapatra:1979ia}. Intriguingly, the baryogenesis and dark matter problems can be solved simultaneously if 
 $M_1\sim $\,keV and  $M_2\sim M_3\sim $\,GeV. 
In fact, in this case $N_{2,3}$ create flavored lepton asymmetries from CP-violating oscillations in the early Universe, which generate the baryon asymmetry of the Universe via ARS leptogenesis \cite{Akhmedov:1998qx}. The lightest sterile neutrino $N_1$ can act as dark matter, with the correct relic abundance achieved through freeze-in production, resonantly enhanced with the MSW effect \cite{Wolfenstein:1977ue,Wolfenstein:1979ni,Mikheev:1986gs}.
Moreover, it was argued in Ref.  \cite{Bezrukov:2007ep} that the puzzle of inflation  can be solved even in the SM by allowing a non-minimal coupling of the Higgs field to the Ricci scalar,  
\begin{equation}
\label{eq:non_min_higgs}
S\supset - \int d^4x\sqrt{- g}\,\xi_H\, H^\dagger H  R,
\end{equation}
which promotes the Higgs field to an inflaton candidate. 

However, the viability of the $\nu$MSM as a minimal model of particle cosmology is threatened by several facts. 
First of all, recent findings in astrophysics have seriously constrained the parameter space for $N_1$ as a dark matter candidate \cite{Schneider:2016uqi,Perez:2016tcq}. Secondly, the generically large value of the non-minimal coupling 
$\xi_H \sim 10^5 \sqrt{\lambda_H}$, where $\lambda_H$ is the Higgs self-coupling, required to fit the amplitude of the scalar perturbations inferred from the 
cosmic microwave background (CMB) temperature fluctuations, imply that perturbative unitarity breaks down at the scale
$M_P/\xi_H\sim 10^{14}$\,GeV, where $M_P=1/\sqrt{8\pi\,G}$ is the reduced Planck mass,
making the inflationary predictions unreliable \cite{Barbon:2009ya,Burgess:2009ea}.  
Even more, successful inflation cannot happen in this context if the quartic coupling $\lambda_H$ in the Higgs potential, 
\begin{equation}
\label{Higgs_potential} 
\nonumber
V(H )  = \lambda_H \left( H^\dagger H - \frac{v^2}{2}\right)^2 ,
\end{equation}
runs negative at large (Planckian) field values due to loop corrections involving the top quark. In fact, 
the central values of the strong gauge coupling and the Higgs and top quark masses imply that $\lambda_H$ becomes negative at a field value corresponding to an energy scale $\Lambda_I \sim 10^{11}$ GeV. This is much lower than what is required for Higgs inflation and thus inconsistent with it. However, given the current experimental uncertainties, a definite conclusion cannot yet be drawn, see e.g.~\cite{Buttazzo:2013uya,Bednyakov:2015sca}. 

These obstacles of the $\nu$MSM can be neatly circumvented in SMASH-type 
\cite{Ballesteros:2016euj,Ballesteros:2016xej,Ernst:2018bib} extensions of the SM which are built around the axion for the solution of the strong CP problem \cite{Peccei:1977hh,Weinberg:1977ma,Wilczek:1977pj}, as well as for dark matter, and allow inflation to be driven by 
(a mixture of the modulus of the Higgs field with) the modulus of the Peccei-Quinn field --sometimes called saxion field  \cite{Pi:1984pv,Fairbairn:2014zta}.    

This review is organized as follows. In Sect. \ref{model} we describe a number of Peccei-Quinn-type extensions of the $\nu$MSM:
bottom-up constructions featuring KSVZ- and DFSZ-type axions (cf. Sects. \ref{sec:SMASH} and \ref{sec:2hdSMASH}, respectively) 
and top-down constructions based on non-supersymmetric grand unification (cf. Sect. \ref{sec:gutSMASH}). Section \ref{sec:inflation} is devoted to inflation, while stability is analyzed in section \ref{sec:stability}. Reheating is reviewed in Sect.~\ref{sec:reheating}, dark matter in Sect.~\ref{sec:dark_matter}, and baryogenesis in Sect.~\ref{sec:baryogenesis}. Conclusions are drawn in Section \ref{sec:conclusions}.

\section{SMASH AND ITS VARIANTS}
\label{model}

In this section we will describe a number of extensions of the SM which exploit the Peccei-Quinn (PQ) mechanism \cite{Peccei:1977hh} to solve the strong CP problem and thus have the potential to solve the big five problems of particle physics and 
cosmology in one smash.

\subsection{SMASH}
\label{sec:SMASH}

\begin{figure}[t]
\begin{center}
\includegraphics[width=0.6\textwidth]{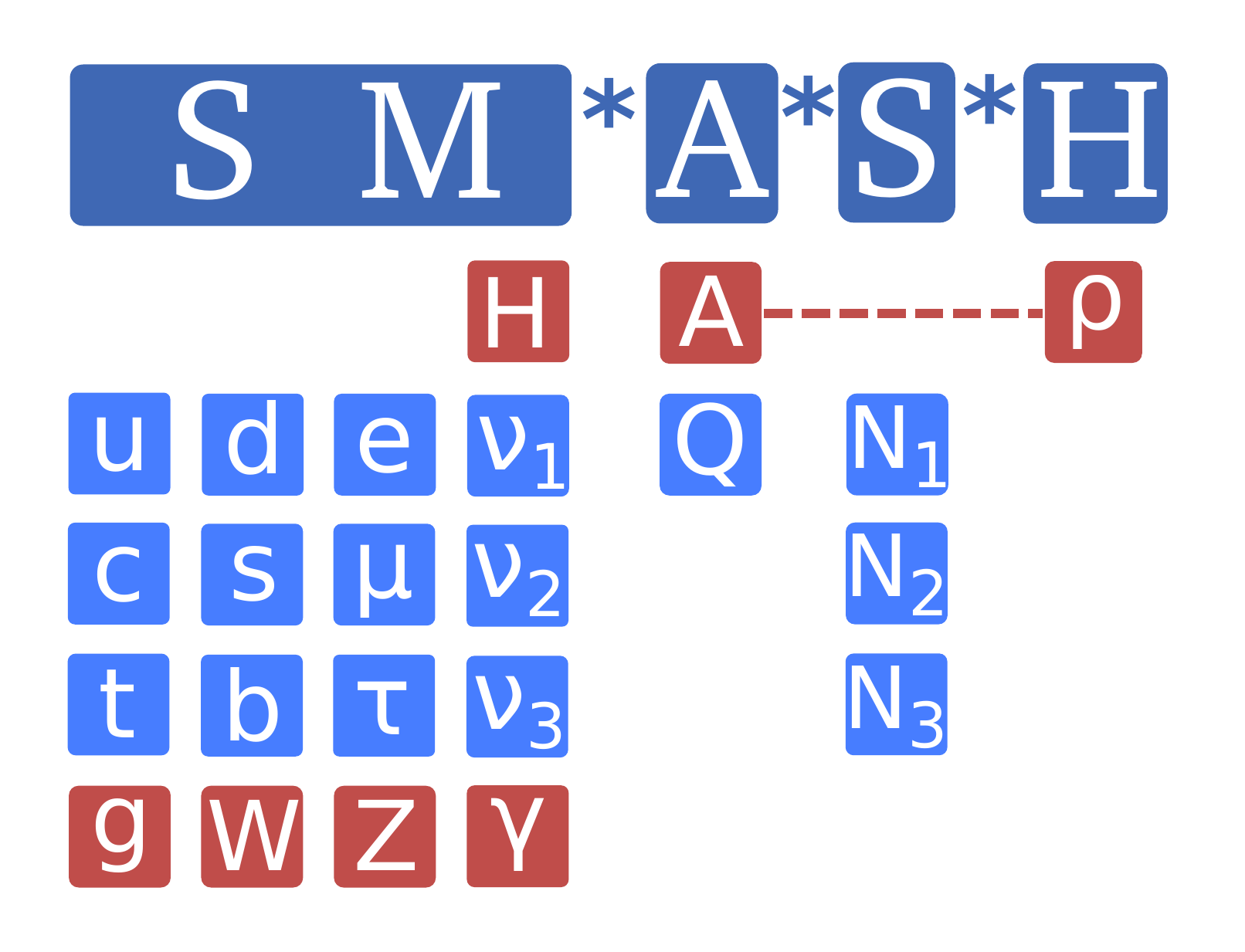}
\caption{\label{fig:SMASH} Particle/field content of SMASH. 
}
\end{center}
\end{figure}

The model with smallest field content -- dubbed here and in the following SMASH -- is based on a KSVZ-type axion model
\cite{Kim:1979if,Shifman:1979if}: a SM-singlet complex scalar field $\sigma$, which features a (spontaneously broken) global $U(1)_{\rm PQ}$ symmetry, and a  
vector-like coloured Dirac fermion $Q$, which transforms as\footnote{These hypercharge assignments ensure that $Q$ can mix with the 
right-handed SM down-type quarks or up-quarks, respectively, allowing its decay to the latter, thereby evading overabundance problems \cite{Nardi:1990ku,Berezhiani:1992rk}.} $(3,1,-1/3)$ or, alternatively, as $(3,1,2/3)$ under the SM
gauge group $SU(3)_C\times SU(2)_L\times U(1)_Y$ and which transforms chirally under $U(1)_{\rm PQ}$, are added to the field content 
of the $\nu$MSM (cf. Fig. \ref{fig:SMASH}). 
The scalar potential, which relates the Higgs field $H$ to $\sigma$, is assumed to have the general form
\begin{equation}
\label{scalar_potential} 
\nonumber
V(H,\sigma )  = \lambda_H \left( H^\dagger H - \frac{v^2}{2}\right)^2
+\lambda_\sigma \left( |\sigma |^2 - \frac{v_{\sigma}^2}{2}\right)^2 +
2\lambda_{H\sigma} \left( H^\dagger H - \frac{v^2}{2}\right) \left( |\sigma |^2 - \frac{v_{\sigma}^2}{2}\right) ,
\end{equation}
with $\lambda_H, \lambda_\sigma >0$ and 
$\lambda_{H\sigma}^2 <  \lambda_H \lambda_\sigma$, in order to ensure that both the electroweak  symmetry and the PQ symmetry are broken in the vacuum; i.e.\ the minimum of the scalar potential
is attained at  the vacuum expectation values (VEVs)  
\begin{equation}
\langle{H^\dagger H}\rangle = v^2/2, \hspace{6ex}
\langle{|\sigma |^2}\rangle=v_{\sigma}^2/2\,,
\end{equation}
where $v=246$\,GeV. 
The PQ symmetry breaking scale $v_\sigma$ is assumed to be much larger than the Higgs VEV $v$.
Correspondingly,  the particle excitation of the modulus  $\rho$ of $\sigma$, cf. 
\begin{equation}
\label{sigma:}
\sigma (x) =\frac{1}{\sqrt{2}}\big[v_{\sigma}+\rho (x)\big]e^{iA(x)/v_{\sigma}}
\,,
\end{equation}
gets a large mass  
\begin{equation}
m_\rho =   \sqrt{2\,\lambda_\sigma}\, v_{\sigma} + \mathcal{O}\left(  \frac{v}{v_{\sigma}}\right)\,,
\end{equation}
while the particle excitation $A$ of the angular degree of freedom of $\sigma$ 
--  which is dubbed ``axion" in the context of the PQ solution of the strong CP problem \cite{Weinberg:1977ma,Wilczek:1977pj} -- is a massless Nambu-Goldstone (NG)
boson, $m_A=0$. 

However, due to the assumed chiral transformation of the new vector-like fermion $Q$, 
the $U(1)_{\rm PQ}$ symmetry is broken due to the gluonic triangle anomaly, 
\begin{equation}
\partial_\mu J^\mu_{U(1)_{\rm PQ}} \supset 
- \frac{\alpha_s}{8\pi}\,G_{\mu\nu}^c {\tilde G}^{c,\mu\nu}\,.
\end{equation} 
Under these circumstances, the NG field  
\begin{equation}
\theta(x) \equiv \frac{A(x)}{f_A}\,,\ {\rm with\ }\ f_A \equiv \frac{v_\sigma}{N_{\rm DW}}\  {\rm \ and\ }\ N_{\rm DW}=1\,,
\end{equation} 
acts as a space-time dependent $\theta$-angle in QCD. In fact, the anomaly ensures that, 
at energies above the scale of QCD, $\Lambda_{\rm QCD}$, but far below the scale of PQ symmetry 
breaking, $v_\sigma$,  that is after integrating out 
the saxion $\rho$ and the vector-like quark $Q$, which also gets a large mass from its Yukawa coupling with the PQ scalar, 
\begin{equation}
m_Q = \frac{y}{\sqrt{2}}\, v_{\sigma} + \mathcal{O}\left(  \frac{v}{v_{\sigma}}\right)\,, 
\end{equation} 
the effective Lagrangian of the axion has the form 
\begin{equation}
{\mathcal L}_\theta = 
\frac{f_A^2}{2} \,\partial_\mu \theta \partial^\mu \theta
- \frac{\alpha_s}{8\pi}\,\theta(x)\,G_{\mu\nu}^c {\tilde G}^{c,\mu\nu}\,.
\end{equation}
Correspondingly, the $\overline\theta$-angle in QCD can be eliminated by a shift $\theta (x) \to \theta (x) -\overline\theta$. 
At energies below $\Lambda_{\rm QCD}$, 
the effective potential of the shifted field, which for convenience we again denote by $\theta(x)$, will then coincide 
with the vacuum energy of QCD as a function of $\overline\theta$,  
 \begin{equation}
\label{axion-potential}
V(\theta ) \equiv 
- \frac{1}{\mathcal V} \ln \frac{Z (\overline\theta)}{Z (0)}\Bigg{|}_{\overline\theta = \theta}
\simeq  \Sigma_0  \, \left( m_u + m_d \right) \left( 1 -  
\frac{\sqrt{m_u^2+m_d^2 + 2 m_u m_d \cos  \theta  }}{m_u + m_d}\right) ,
\end{equation}
where $\mathcal{V}$ is the Euclidean space-time volume, $Z (\overline\theta)$ is the partition function of QCD, and
$\Sigma_0 = -\langle \bar u u\rangle = -\langle \bar d d\rangle$
is the chiral condensate  \cite{DiVecchia:1980yfw,Leutwyler:1992yt}. 
Notably, CP is conserved in the vacuum, since 
$V(\theta )$ has an absolute minimum at $\theta=0$ and thus 
the vacuum expectation value of $\theta$ vanishes, $\langle \theta\rangle =0$  \cite{Vafa:1984xg}. 
Expanding the potential around zero and using
\begin{equation}
m_\pi^2 = \frac{\Sigma_0}{f_\pi^2} (m_u + m_d) + {\mathcal O}(m^2),
\end{equation}
one finds the mass of the axion as the coefficient of the quadratic term, 
\begin{equation}
\label{axion-mass}
m_A \equiv \frac{\sqrt{\chi_0}}{f_A}\simeq \frac{m_\pi f_\pi}{f_A} \frac{\sqrt{m_u m_d}}{m_u+m_d},
\end{equation}
where 
$\chi_0$ 
is the topological susceptibility in QCD, $m_\pi = 135$ MeV the neutral pion mass, $f_\pi \approx 92$ MeV its decay constant, and $m_u$, $m_d$ are the masses of the lightest quarks, with 
ratio $z=m_u/m_d \approx 0.56$. A recent determination in next-to-leading order chiral perturbation theory  \cite{diCortona:2015ldu} yielded 
$\chi_0 = [75.5(5) {\rm MeV}]^4$,
which agrees beautifully with the result from lattice QCD, 
$\chi_0=[75.6(1.8)(0.9) {\rm MeV}]^4$ \cite{Borsanyi:2016ksw},
resulting in 
\begin{equation}
\label{zeroTma}
m_A= 
{57.0(7)\,   \left(\frac{10^{11}\rm GeV}{f_A}\right)\mu \textrm{eV}. }
\end{equation} 

Moreover, also couplings to the photon and the nuclei are inherited from the axion's mixing with the pion. 
The full low energy Lagrangian of the axion with photons ($F_{\mu\nu}$), nucleons, $\psi_N=p,n$, electrons ($e$) 
and active neutrinos ($\nu_i$) has the generic form   
\begin{eqnarray}
\label{axion_leff_app}
{\mathcal L}_A &= &\frac{1}{2} \partial_\mu A \partial^\mu A - V(A) 
- \frac{\alpha}{8\pi}\,C_{A\gamma}\frac{A}{f_A}\,F_{\mu\nu} {\tilde F}^{\mu\nu} 
+ \frac{1}{2}  C_{AN}\frac{\partial_\mu A}{f_A} \ \overline\psi_N \gamma^\mu\gamma_5 \psi_N
\\ \nonumber
&&
+ \frac{1}{2}  C_{Ae}\frac{\partial_\mu A}{f_A} \ \overline\psi_e \gamma^\mu\gamma_5 \psi_e
+ \frac{1}{2} C_{A\nu}  \frac{\partial_\mu A}{f_A} \ \overline\nu_i \gamma^\mu\gamma_5 \nu_i
\,,
\end{eqnarray}
where $V(A)=V(\theta = A/f_A)$. The dimensionless coupling to photons,  $C_{A\gamma}$, involves a model-independent part from the mixing with the pion and a model-dependent part depending of the electric charge of $Q$. It is given in 
Table \ref{tab:couplings_ksvz} for the two variants of SMASH. 
Similarly, the proton and neutron have a model-independent part and
a model dependent contribution that arises from possible axion-quark couplings
of the form $(C_{Aq}/2)(\partial_\mu A/f_A)\bar \psi_q \gamma^\mu\gamma_5\psi_q$
in the high-energy theory,
\begin{align}
\label{eq:nucleon_couplings}
C_{Ap} & = -0.47(3)+0.88(3) C_{Au} -0.39(2) C_{Ad}-0.038(5)C_{As} \nonumber \\
				&\hspace{2.15cm}-0.012(5) C_{Ac} -0.009(2) C_{Ab}-0.0035(4) C_{At}\,, \nonumber  \\
C_{An} & = -0.02(3)+0.88(3) C_{Ad} -0.39(2) C_{Au}-0.038(5)C_{As} \nonumber \\
				&\hspace{2.15cm}-0.012(5) C_{Ac} -0.009(2) C_{Ab}-0.0035(4) C_{At} \,,
\end{align}
as found in the state-of-the-art calculation~\cite{diCortona:2015ldu}. In SMASH, all the axion-quark and axion-charged-lepton 
couplings vanish at tree level, cf. Table \ref{tab:couplings_ksvz}. 
\begin{table}[t]
$$
\begin{array}{|c|c||c|c|c|c|}
\hline
 \bf Model &R_Q & f_A  & N_{\rm DW} &C_{A\gamma} & C_{Ai}   \\
\hline
\rm SMASH(d)&(3,1,-\tfrac{1}{3})	 & v_\sigma & 1 & \frac{2}{3} - 1.92(4)& 0    \\
\hline
\rm SMASH(u)& (3,1,+\tfrac{2}{3})  & v_\sigma &  1& \frac{8}{3} - 1.92(4)& 0    \\
\hline
\end{array}
$$
\caption{\label{tab:couplings_ksvz}
Axion predictions for two SMASH variants exploiting distinct vector-like quarks transforming as $R_Q$ under the SM
gauge group factors $SU(3)_C\times SU(2)_L\times U(1)_Y$: Axion decay constant $f_A$, domain wall number $N_{\rm DW}$,  coupling to the photon 
$C_{A\gamma}$, and tree-level couplings to quarks and charged leptons $C_{Ai}$, $i=u,...,t,e,..,\tau$. 
}
\end{table}
%

To avoid strong bounds from laboratory experiments and stellar astrophysics, the axion decay constant $f_A$ has 
to be  much larger than the electroweak scale \cite{Tanabashi:2018oca}, 
notably $f_A \gtrsim10^8\ {\rm GeV}$ from the measured duration of the neutrino signal of supernova 1987A \cite{Raffelt:2006cw,Fischer:2016cyd,Chang:2018rso}.

Optionally, one may unify the PQ symmetry with a lepton number symmetry  by assigning PQ charges also to the leptons and sterile neutrinos \cite{Shin:1987xc,Dias:2014osa} .
In this case, 
the latter get their Majorana masses also from PQ symmetry breaking, 
\begin{equation}
M_{ij} = \frac{Y_{ij}}{\sqrt{2}} v_\sigma + \mathcal{O}\left(  \frac{v}{v_{\sigma}}\right)\,,
\end{equation}
where $Y_{ij}$ are Yukawa couplings,  
and the mass scale of the active neutrinos is determined by the PQ scale, 
\begin{equation} 
m_{\nu} =  0.04\,{\rm eV}  \left( \frac{10^{11}\,{\rm GeV}}{v_{\sigma}} \right)
\left( \frac{-  F\,Y^{-1}\,F^T}{10^{-4}}\right)\,.
\end{equation} 
Moreover, the axion $A$ is in this case at the same time the majoron $J$: the NG boson arising from the breaking
of the global lepton number symmetry \cite{Chikashige:1980ui,Gelmini:1980re,Schechter:1981cv}. This 
leads to  a non-zero tree-level coupling of the $A/J$ to the active neutrinos, $(-1/4)(\partial_\mu A/f_A)\bar \nu_i \gamma^\mu\gamma_5\nu_i$ and to possibly sizeable loop-induced couplings to SM quarks and charged leptons from the loop involving the sterile neutrinos $N_{i}$ \cite{Shin:1987xc,Pilaftsis:1993af}. To lowest order in the seesaw limit,  $m_D/M_M\ll 1$, they are given by \cite{Garcia-Cely:2017oco}
\begin{align}
\label{CaiAXIMAJ}
C_{aq} \simeq \frac{1}{8\pi^2}  \, T^{q}_3\, {\rm tr} \kappa \,, \hspace{3ex}
C_{A\ell} \simeq -\frac{1}{16\pi^2} \left(  {\rm tr} \kappa -2 \kappa_{\ell\ell}   \right)
 \,,
\end{align}
where $T^{d}_3 = -\tfrac12 = - T^u_3$ and the dimensionless hermitian $3\times 3$ matrix $\kappa$ is defined as
\begin{align}
\kappa \equiv  \frac{m_D m_D^\dagger}{v^2} = \frac{F F^\dagger}{2}\, .
\label{eq:K}
\end{align}
Intriguingly, a KSVZ-type axion/majoron with $f_A\sim 10^8$\,GeV may explain the $\sim 3\,\sigma$ hint of an 
anomalously large energy loss of helium burning stars, red giants and white dwarfs, 
if $|\kappa - 2\kappa_{ee}|$ is of order unity  \cite{Giannotti:2017hny}.

\subsection{2hdSMASH}
\label{sec:2hdSMASH}

A less minimal variant of SMASH -- dubbed 2hdSMASH -- exploits 
DFSZ-type axion models~\cite{Zhitnitsky:1980tq,Dine:1981rt}:  in those the SM Higgs sector is extended by 
two Higgs doublets, $H_u$ and $H_d$, whose vacuum expectation values $v_u$ and $v_d$ give masses to up-type
and down-type quarks, respectively. There are two possibilities,  named 2hdSMASH(d) or 2hdSMASH(u), 
according to whether leptons couple to $H_d$, which occurs in familiar Grand Unified Theories (GUTs), or to $H_u$. The SM model quarks are assumed to carry PQ charges such that the gluonic triangle anomaly arises from them alone. The low-energy Lagrangian of a DFSZ-type PQ extension of the SM is identical to that of a 2 Higgs Doublet Model (2HDM), augmented by seesaw-generated neutrino masses Eq. \eqref{eq:seesaw_neutrino_masses}, and the one of a DFSZ-type axion.  
The DFSZ axion properties are given in  Table \ref{tab:couplings_dfsz}. In this case, there are tree-level couplings to quarks and leptons. 
In fact, the anomalous stellar energy losses mentioned above can be alternatively explained
by a DFSZ-type  axion with $f_A\gtrsim 10^8$\,GeV and $\tan\beta\equiv v_u/v_d \sim 1$ \cite{Giannotti:2017hny}.

\begin{table}[h]
$$
\begin{array}{|c||c|c|c|c|c|c|}
\hline
\bf Model & f_A & N_{\rm DW} & C_{A\gamma}  & C_{Au} & C_{Ad}  & C_{A\ell}   \\
\hline
 \rm 2hdSMASH(d) & v_\sigma/6 & 6 & \frac{8}{3} - 1.92(4)& \frac{1}{3} \cos^2 \beta & \frac{1}{3} \sin^2 \beta & \frac{1}{3} \sin^2 \beta \\
\hline
 \rm 2hdSMASH(u) & v_\sigma/6 & 6 &\frac{2}{3} - 1.92(4)& \frac{1}{3} \cos^2 \beta & \frac{1}{3} \sin^2 \beta  & \frac{1}{3} \cos^2 \beta
\\[.5ex]
\hline
\end{array}
$$
\caption{\label{tab:couplings_dfsz}
DFSZ-type axion predictions: Axion decay constant $f_A$, domain wall number $N_{\rm DW}$,  coupling to the photon 
$C_{A\gamma}$, and tree-level couplings to quarks and charged leptons $C_{Ai}$, $i=u,...,t,e,..,\tau$, with $\tan\beta \equiv v_u/v_d$.  
}
\end{table}

Again, optionally the PQ symmetry may be unified with a lepton number symmetry \cite{Langacker:1986rj,Volkas:1988cm,Clarke:2015bea}, in which case the active neutrino mass scale is determined by the PQ scale and the DFSZ axion is at the same time a Majoron.

\subsection{gutSMASH}
\label{sec:gutSMASH}

As commented in the previous section, the model 2hdSMASH(d) can be embedded into a GUT. The simplest unified group is $SU(5)$ \cite{Georgi:1974sy,Georgi:1974my}, with each generation of fermions
(not including right-handed neutrinos) fitting into the representations $10_F$ and  $\bar{5}_F$, with $SU(5)$ broken into the SM group by the VEV of a scalar in the $24_H$, and with the electroweak breaking carried out by two scalars in the $5_H$. It was realized early on that $SU(5)$ GUTs can accommodate an axion with a decay constant $f_A$ tied to the unification scale  \cite{Wise:1981ry}. However, minimal nonsupersymmetric $SU(5)$ GUTs are incompatible with proton decay limits, because the $SU(2)$ and $U(1)$ gauge couplings meet at too low a scale. However, there are viable extensions in which particles in additional $SU(5)$ multiplets appropriately modify the running of the gauge couplings so as to yield successful unification compatible with proton decay limits. The extension proposed in \cite{Bajc:2006ia} and further studied in  \cite{Bajc:2007zf,DiLuzio:2013dda} makes use of a fermionic multiplet in the $24_F$, which contains right-handed neutrinos getting a mass from the VEV of the $24_H$, which breaks $SU(5)$ into the SM. This generates masses for the light neutrinos through a combination of the type I and III seesaw mechanisms, and also allows for baryogenesis from leptogenesis. When extending this viable $SU(5)$ model to accommodate a global PQ symmetry with its corresponding axion \cite{DiLuzio:2018gqe}, one has a SMASH-type construction with the complex scalar in the $24_H$ containing the axion and acting as a Majoron. The Lagrangian  of this model, which we will refer to as miniSU(5)PQ, contains the following interactions (written only schematically),
\begin{align}
{\mathcal L}\supset\bar 5_F 10_F 5'^*_H + 10_F 10_F 5_H + \bar 5_F 24_F 5_H + 
{\rm Tr} 24^2_F 24^*_H +5'^*_H 24^2_H 5_H + 5'^*_H 5_H {\rm Tr} (24^2_H)  + \text{h.c.},
\end{align}
which enforce the PQ charge assignments in Table \ref{tab:SU5PQassignments}.
\begin{table}[t]
\begin{align*}
\begin{array}{|c||c|c|c|c|c|c|c|}
\hline
\bf Model & \overline{5}_F & 10_F & 24_F & 5_H & 5'_H & 24_H  \\
\hline
\rm miniSU(5)PQ & 1 & 1 & 1 &-2 & 2 & 2\\
\hline
\end{array}
 \end{align*}
 \caption{\label{tab:SU5PQassignments}Field content and PQ charge assignments in the PQ-extended $SU(5)$ model of \cite{DiLuzio:2018gqe}.}
\end{table}
The axion decay constant is related to the unification scale $v_U$ as $f_A=v_U/11$, while the axion couplings to nucleons and leptons are given in Table \ref{tab:couplings_so10_gut}. 
\begin{table}[b]
$$
\begin{array}{|c||c|c|c|c|c|c|}
\hline
\bf Model & f_A & N_{\rm DW} &  C_{A\gamma}  & C_{Au} & C_{Ad}  & C_{A\ell}   \\
\hline
 \rm miniSU(5)PQ  & v_{\rm U}/11 & 11 & \frac{8}{3} - 1.92(4)& \frac{2}{11} \cos^2 \beta & \frac{2}{11} \sin^2 \beta & \frac{2}{11} \sin^2 \beta \\
\hline
 \rm miniSO(10)PQ  & v_{\rm U}/3 & 3 & \frac{8}{3} - 1.92(4)& \frac{1}{3} \cos^2 \beta & \frac{1}{3} \sin^2 \beta & \frac{1}{3} \sin^2 \beta \\
\hline
 \rm gutSMASH & v_\sigma/3 & 3 &\frac{8}{3} - 1.92(4)& \frac{1}{3} \cos^2 \beta & \frac{1}{3} \sin^2 \beta  & \frac{1}{3} \sin^2 \beta 
\\[.5ex]
\hline
\end{array}
$$
\caption{\label{tab:couplings_so10_gut}
Axion predictions in $SU(5)\times U(1)_{\rm PQ}$ \cite{DiLuzio:2018gqe} and $SO(10)\times U(1)_{\rm PQ}$ models \cite{Ernst:2018bib}: Axion decay constant $f_A$, domain wall number $N_{\rm DW}$,  coupling to the photon 
$C_{A\gamma}$, and tree-level couplings to quarks and charged leptons $C_{Ai}$, $i=u,...,t,e,..,\tau$. In the $SU(5)$ theory, $\tan\beta=v_H/v_H'$, while for the $SO(10)$ models $\tan^2\beta = ((v_u^{10})^2+(v_u^{126})^2)/((v_d^{10})^2+(v_d^{126})^2)$.  
}
\end{table}
The unification scale turns out to be highly constrained and grows with decreasing mass of the light fermion triplet contained in $24_F$. This is due to the fact that increasing the unification scale requires a larger deviation  in the running of the $SU(2)$ and $U(1)$ gauge couplings with respect to the SM case, which can only achieved if the extra particles with electroweak charges in the $24_F$ multiplet become lighter. The light electroweak triplets  can be probed by LHC searches \cite{Arhrib:2009mz,Sirunyan:2017qkz}, which then give upper bounds for $v_U\propto f_A$. On the other hand,  proton decay experiments such as Super-Kamiokande  \cite{Miura:2016krn} constrain the unification scale from below. Given the relation \eqref{zeroTma} between $f_A$ and the axion mass, this results in 
a remarkably constrained window of allowed values of $m_A$:
\begin{equation}
\label{eq:preferred_window}
m_A \in [4.8, 6.6]\  \text{neV}\,.
\end{equation}
The upper limit can be relaxed to $m_A < 330$ neV when allowing for fine-tuning in the flavour structure of the model so as to close as many decay channels for the proton as possible \cite{Dorsner:2004xa}. The above axion mass window can be targeted in a complementary manner by future high-energy colliders \cite{Ruiz:2015zca,Cai:2017mow}, proton decay experiments such as Hyper-Kamiokande \cite{Abe:2011ts}, as well as direct axion dark matter searches with CASPER-Electric \cite{Budker:2013hfa,JacksonKimball:2017elr} and ABRACADABRA \cite{Kahn:2016aff}.

The smallness of the axion mass in this model implies that the axion can be identified with dark matter only if the Peccei-Quinn symmetry is broken before or during inflation  and 
not restored afterwards, as reviewed in  Sect. \ref{sec:dark_matter}. On the other hand, the large value of $f_A$ implies that inflation can source large axionic isocurvature fluctuations which may be in conflict with observations, cf. Sect. \ref{sec:dark_matter}.

Compared to $SU(5)$ GUTs, theories based on the $SO(10)$ group \cite{Georgi:1974my,Fritzsch:1974nn} can yield viable unification patterns without the need to either consider supersymmetric extensions or to add additional fermion multiplets beyond those containing the SM fermions. Moreover, right-handed neutrinos are automatically incorporated, as these  occur automatically with the rest of the SM quarks and leptons if one considers three spinorial representations $16_F$ 
of  $SO(10)$.  The latter can have the following Yukawa couplings with scalar Higgses in the $10_H$ and $\overline{126}_H$ representations,
\begin{equation}
\label{SO10Yukawacomplex}
\mathcal{L}_Y = 16_F \left( Y_{10} 10_H + \tilde{Y}_{10} 10_H^\ast  + Y_{126} \overline{126}_H \right) 16_F + {\rm h.c.} \, ,
\end{equation}
which can give rise to the seesaw mechanism~\cite{GellMann:1980vs}. Moreover, a PQ symmetry, under which the fields transform as 
\begin{align}
\label{pgsymmetry_yukawas}
  16_F\rightarrow16_F e^{i\alpha}\,;\hspace{3ex}
  10_H\rightarrow10_H e^{-2i\alpha}\,;\hspace{3ex}
  \overline{126}_H&\rightarrow\overline{126}_H e^{-2i\alpha}\,,
\end{align} 
can be motivated independently from the strong CP problem:  it forbids
the second term in the Yukawa interactions \eqref{SO10Yukawacomplex}, thereby crucially improving the economy and predictivity of the models \cite{Bajc:2005zf,Babu:1992ia}.

Adding a further Higgs representation, say $210_H$, 
the $SO(10)$ symmetry can be broken at the unification scale $M_{\rm U}$ by the VEV of the $210_H$ to the 
Pati-Salam gauge group $SU(4)_C\times SU(2)_L\times SU(2)_R$, which is broken to the SM gauge group
 $SU(3)_C\times SU(2)_L\times U(1)_Y$ at the scale of $B-L$ breaking $M_{\rm BL}$ (which is thus the seesaw scale) by the VEV of the $\overline{126}_H$, which itself is broken at the weak scale $M_Z$ by the VEV of the $10_H$,  
\begin{eqnarray*} 
SO(10)&\stackrel{M_{\rm U}-210_H}{\longrightarrow}
&4_{C}\, 2_{L}\, 2_{R}\ \stackrel{M_{\rm BL}-\overline{126}_H}{\longrightarrow}3_{C}\, 2_{L}\, 1_{Y}\ \stackrel{M_Z-10_H}{\longrightarrow} \ 3_{C}\,1_{\rm em}\,.
\end{eqnarray*}

Unfortunately, the minimal PQ symmetry \eqref{pgsymmetry_yukawas} leads to a decay constant $f_A= v/3$ \cite{Ernst:2018bib,Mohapatra:1982tc,Holman:1982tb,Altarelli:2013aqa}, which is clearly experimentally excluded. 
The simplest way to remedy this problem is to associate a PQ charge also to the $210_H$, 
\begin{align}
\label{pgsymmetry_GUT}
  210_H\rightarrow 210_H e^{4i\alpha}\,.
\end{align} 

We dub this model miniSO(10)PQ -- for Minimal $SO(10)\times U(1)_{\rm PQ}$ model -- and summarize the field content and 
PQ charge assignments in the first row of Table \ref{tab:PQassignments}. 
Its axion properties are given in Table \ref{tab:couplings_so10_gut}. 
\begin{table}[t]
\begin{align*}
\begin{array}{|c||c|c|c|c|c|}
\hline
\bf Model & 16_F & \overline{126}_H & 10_H & 210_H  & \sigma   \\
\hline
\rm miniSO(10)PQ  & 1 & -2 &-2 & 4   & -   \\
\hline
 \rm gutSMASH & 1 & -2 &-2 & 0  & 4   \\
\hline
\end{array}
 \end{align*}
 \caption{\label{tab:PQassignments}Field content and PQ charge assignments in two distinct $SO(10)\times U(1)_{\rm PQ}$ models \cite{Ernst:2018bib}.}
\end{table}
The photon and fermion couplings are the same as for 2hdSMASH(d),  although the 
microscopic origin of the parameter $\beta$ differs, as it is determined by the VEVs of four Higgses, as opposed to two in DFSZ models. 
Moreover, as in miniSU(5)PQ, the decay constant in miniSO(10)PQ is proportional to the scale of grand unification, $f_A= v_{\rm U}/3$, 
which is determined by the requirement of gauge coupling unification. Therefore, this model is more predictive in the axion sector than 
SMASH or 2hdSMASH, yet less predictive than miniSU(5)PQ due to the additional freedom inherent in having a multi-step breaking of the grand unified group --as opposed to the single-step breaking in the SU(5) case-- as well as due to the additional threshold corrections that can arise from the greater number of particles included in the $SO(10)$ multiplets. Allowing for a reasonable range of scalar threshold corrections and taking into account constraints from 
black hole superradiance \cite{Arvanitaki:2014wva} and proton decay, the axion decay constant and mass is predicted to lie in the range
\cite{Ernst:2018bib}
\begin{equation} \label{fgut}
2.6 \times 10^{15} {\rm GeV} <f_A<3.0\times 10^{17}\text{GeV}, \hspace{3ex}
1.9 \times 10^{-11}{\rm eV} < m_A<2.2\times 10^{-9} {\rm eV}. 
\end{equation} 
As in the miniSU(5)PQ model, such light axion can only be compatible with dark matter with a pre-inflationary breaking of the PQ symmetry, and isocurvature constraints can be important. In fact, a one-step breaking model analogous to miniSU(5)PQ can also be realized in $SO(10)$ by breaking the group at a high scale not just with the $210_H$, but with the added effect of a nonzero  VEV in a $45_H$ scalar multiplet \cite{Boucenna:2018wjc}. In this model, successful unification with a proton lifetime in reach of Hyper-Kamiokande is achieved by ensuring that the octets and triplets inside the $210_H$ remain light, in analogy with the light triplets in miniSU(5)PQ.  The PQ charge of the $210_H$ is now zero, while the $45_H$ is assigned charge 4, which still gives a GUT-scale axion  with a low mass and thus affected by isocurvature constraints.

Such constraints can be definitely evaded in the $SO(10)\times U(1)_{\rm PQ}$ variant dubbed gutSMASH whose field 
content and PQ charge assignments are specified in the second row of Table \ref{tab:PQassignments}. In this model the $210_H$ 
has no PQ charge. Instead, it features a further complex singlet scalar $\sigma$ which is charged under the PQ symmetry. Its 
VEV determines the PQ symmetry breaking scale (see also \cite{Babu:2015bna,Boucenna:2017fna}) and the  
axion decay constant turns out to be $f_A=  v_\sigma/3$ \cite{Ernst:2018bib} (cf. second row of Table \ref{tab:couplings_so10_gut}),
which is a free parameter of the model.

\section{INFLATION}
\label{sec:inflation}

In SMASH and its variants, introduced in the last section, there are two or more scalar fields that in principle could have driven primordial inflation. Let us look into this issue in some detail.

In SMASH, the modulus of the complex PQ field,  $\rho^2=2\,|\sigma|^2$, or a mixture of it with
$h$, the neutral component of the Higgs doublet in the unitary gauge, $H^t=(0\,,h)/\sqrt{2}$, is a viable inflaton candidate. It was pointed out in \cite{Bezrukov:2007ep} that a non-minimal coupling of the Higgs, $H$, to the Ricci scalar $R$, cf. Eq. \eqref{eq:non_min_higgs},  
would allow $h$ to play that role, in a model that is since dubbed Higgs inflation. Indeed,  after scalar and metric field redefinitions into the so-called Einstein frame, this kind of coupling flattens any quartic potential, making it convex and asymptotically flat at large field values \cite{Salopek:1988qh}, approaching a plateau-like form which is preferred by CMB measurements \cite{Akrami:2018odb}. 
However, as mentioned in the Introduction, a large value of the non-minimal coupling
$\xi_H$ --as required to fit the amplitude of the primordial scalar fluctuations ($\xi_H\sim 5\times 10^4 \sqrt{\lambda_H}$) for the central value of the top quark mass \cite{Tanabashi:2018oca} (see also Fig.\ 14 of \cite{Aaboud:2018zbu})--  implies that perturbative 
unitarity breaks down at a scale $M_P/\xi_H$, well below the Higgs field values during inflation inflation $h\sim M_P/\sqrt{\xi_H}$ and comparable to the scale given by the fourth square root of the potential \cite{Barbon:2009ya,Burgess:2009ea}. See \cite{Hamada:2014iga, Bezrukov:2014bra} for the statistically disfavored possibility of reducing $\xi_H$ by considering significantly smaller top masses.

\begin{figure}
\begin{center}
\includegraphics[width=0.3\textwidth]{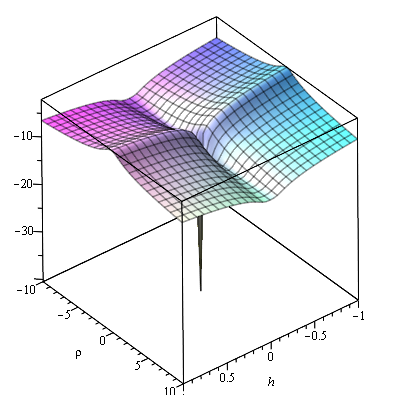}
\includegraphics[width=0.3\textwidth]{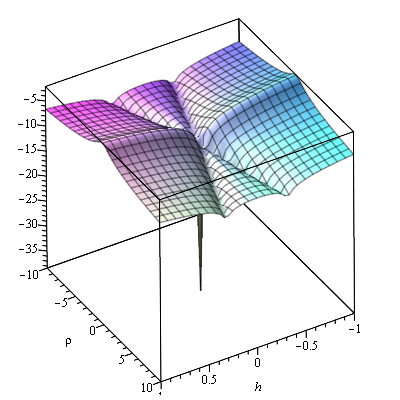}
\caption{Decadic log of the scalar potential \eqref{scalar_potential} in the Einstein frame ($\xi_H\ll \xi_\sigma$), as a function of $h$ and $\rho$, all in units of $M_P$, supporting, 
for $\lambda_{H\sigma}>0$, pure Hidden Scalar Inflation (HSI)
({\em left panel}),
and, for $\lambda_{H\sigma}<0$,  Higgs-Hidden Scalar Inflation (HHSI) ({\em right panel}) \cite{Ballesteros:2016xej}. Inflation proceeds along one of the valleys.
The couplings have been chosen such that the amplitude of primordial scalar perturbation is properly normalised.
}
\label{fig:pot_einstein}       
\end{center}
\end{figure}

\begin{figure}
\begin{center}
\includegraphics[width=6.7cm]{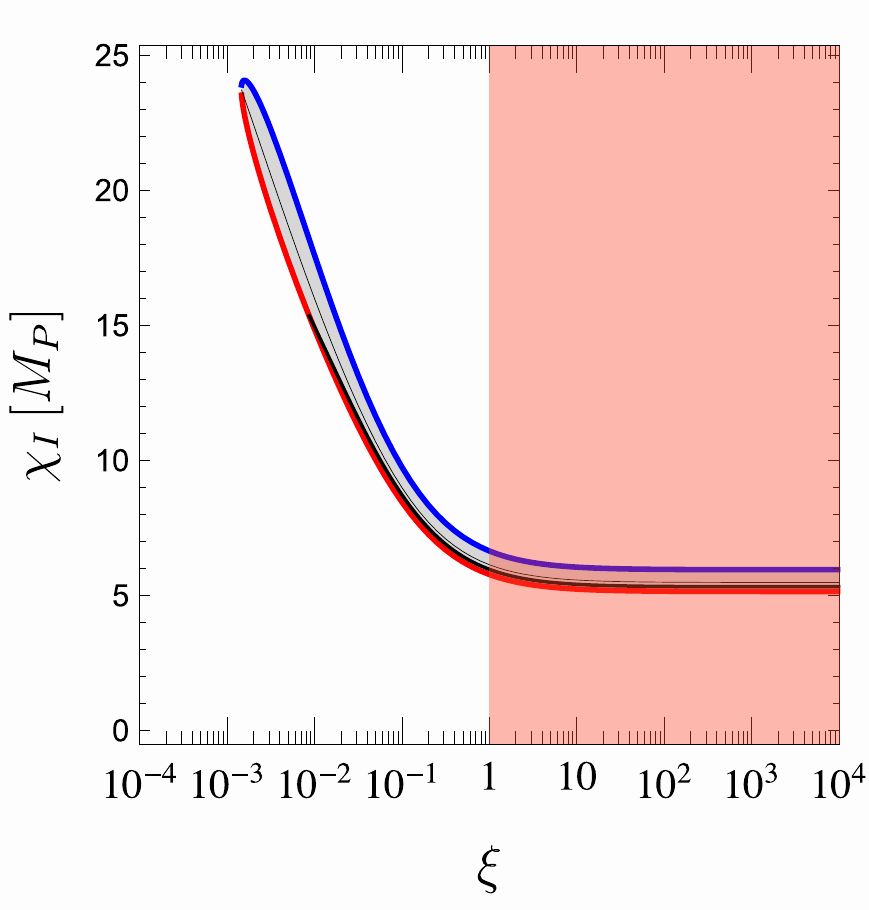} \includegraphics[width=7cm]{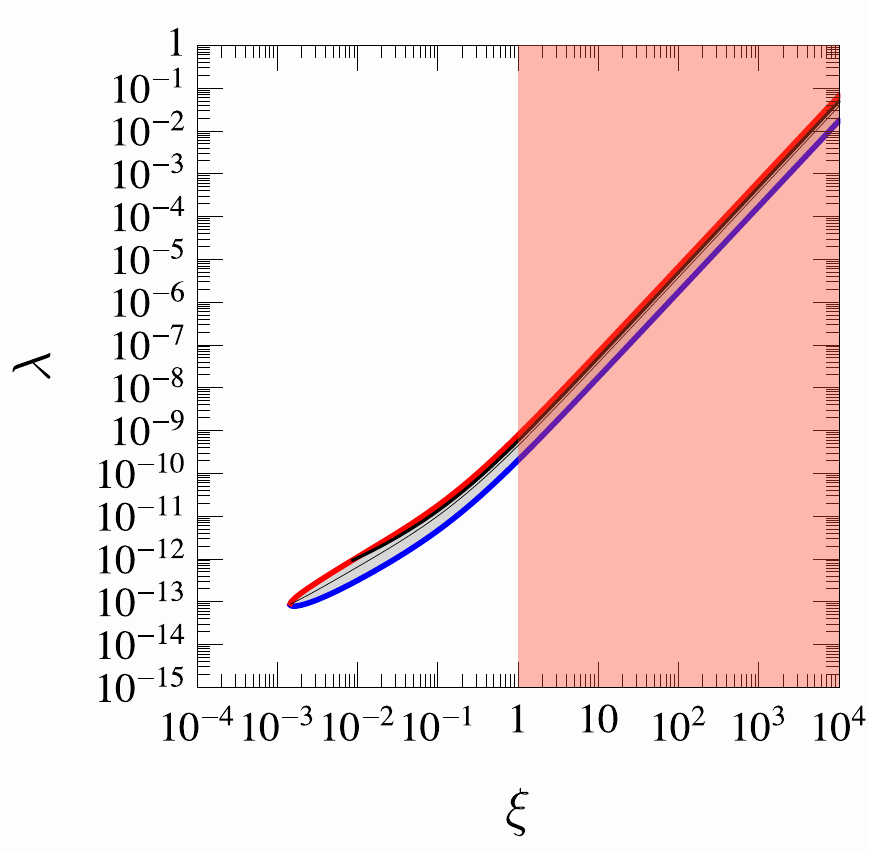}
\includegraphics[width=7.cm]{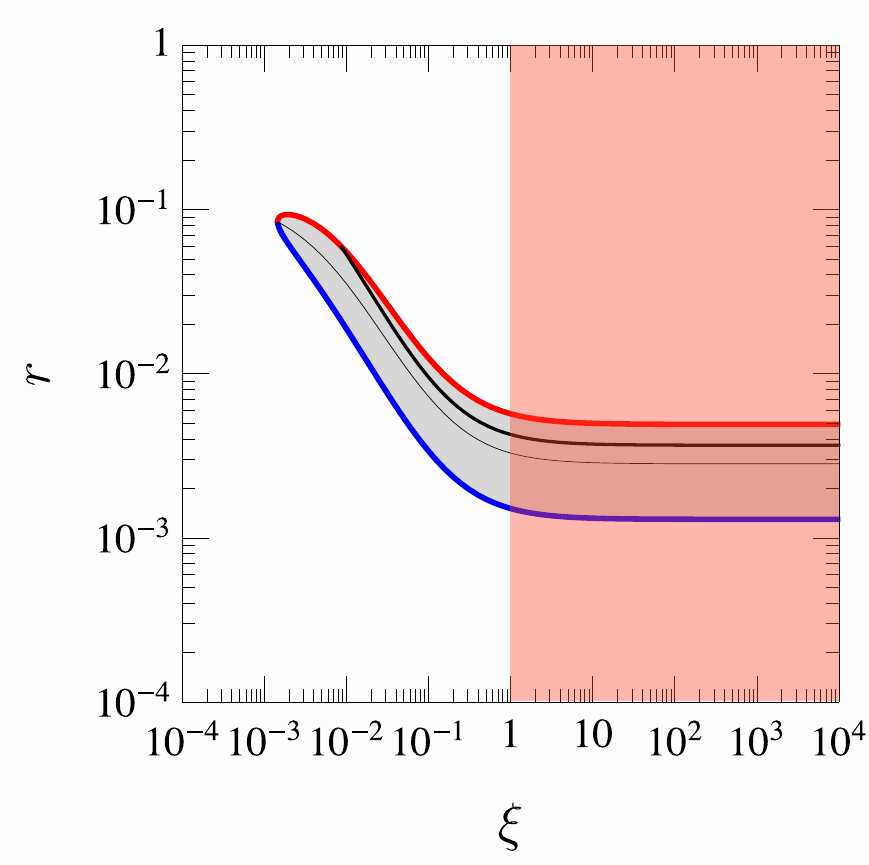} \includegraphics[width=7.1cm]{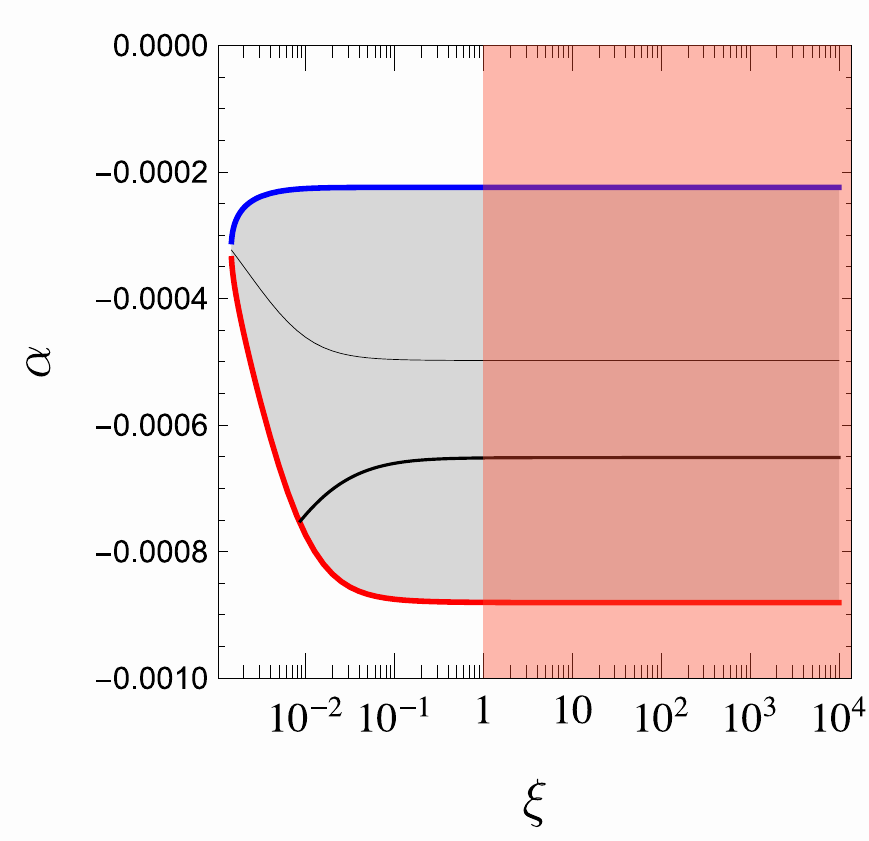}
\caption{95\% C. L. contours for the parameters of the non-minimally coupled potential \eqref{genpotential} giving inflation as constrained by cosmological observations (Planck 2015) at the pivot scale $0.05$ Mpc$^{-1}$ \cite{Ballesteros:2016xej}. Shown are: the canonical inflaton value $\chi_I$ (up left), the value of the quartic coupling (up right), the predicted tensor-to-scalar ratio (down left) and  the predicted running of the spectral index (down right), as a function of the non-minimal coupling parameter $\xi=\xi_\sigma$. The thin black line corresponds to the best fit for a given $\xi_\sigma$, while the red and blue curves correspond to minimum and maximum values of $n_s$, i.e. to a redder or bluer primordial spectrum of curvature perturbations. The thicker black line corresponds to the predictions when taking into account the HHSI prediction of radiation domination immediately after inflation. The shaded regions for $\xi_\sigma>1$ indicate, approximately, the region where the predictivity of inflation is threatened by the breakdown of perturbative unitarity.}
\label{fig:chi2}
\end{center}
\end{figure}

This problem can be eliminated in Hidden Scalar Inflation (HSI) \cite{Pi:1984pv,Fairbairn:2014zta,Boucenna:2017fna} or Higgs-Hidden Scalar  inflation (HHSI) 
\cite{Ballesteros:2016euj,Ballesteros:2016xej}, which exploit a  non-minimal coupling analogous to the previous one:
\begin{equation}
S\supset -\int d^4x \sqrt{-g}\,\xi_\sigma\,\sigma^* \sigma\,R\,.
\end{equation} 
Such couplings are not ad-hoc, since they are generated radiatively in a Friedman-Robertson-Walker background. For negligible $\xi_H$, slow-roll inflation with a tree-level asymptotically flat potential can thus happen along two different directions in field space: the  $\rho$-direction for $\lambda_{H\sigma}>0$ (HSI) and the {line} $h/\rho=\sqrt{-\lambda_{H\sigma}/\lambda_H}$ for $\lambda_{H\sigma}<0$ (HHSI),
cf. Fig.  \ref{fig:pot_einstein}. In both cases, inflation can be described {in the Einstein frame}  by a single canonically normalised field $\chi$ with potential
\begin{equation}
\label{genpotential}
\tilde V(\chi) = \frac{\lambda }{4}\rho(\chi)^4\left(1+\xi_{\sigma}\frac{\rho(\chi)^2}{M_P^2}\right)^{-2}\,,
\end{equation} 
where 
\begin{equation}
\label{effective_quartic_couplings}
{\lambda} \equiv 
\left\{
\begin{array}{ll} 
\lambda_\sigma
,  & \mathrm{for\ HSI},  \\ 
\lambda_\sigma \left( 1-\frac{\lambda_{H\sigma}^2}{\lambda_\sigma\lambda_H} \right)
,  & \mathrm{for\ HHSI}\,.
\end{array}
\right.
\end{equation}
The field $\chi$ is the solution of $\Omega^2\,d\chi/d\rho\simeq (b\,\Omega^2+6\,\xi_\sigma^2\,\rho^2/M_P^2)^{1/2}$, with $\Omega\simeq 1+\xi_\sigma\,\rho^2/M_P^2$ being the Weyl transformation into the Einstein frame and 
{$b=1$ (for HSI) or $b=1+|\lambda_{H\sigma}/\lambda_H|$ (for HHSI). We will see in the next section that vacuum stability requires a small value of 
$|\lambda_{H\sigma}|\lesssim 10^{-6}$ and consequently $b\sim1$ in HHSI, which makes practically impossible distinguishing between HSI and HHSI from the measurements of the CMB power spectra. However, even a small Higgs component in the inflaton is a key aspect for reheating, which sets apart both possibilities, as we will discuss later.

\begin{figure}[t]
\begin{center}
\includegraphics[width=0.9\textwidth]{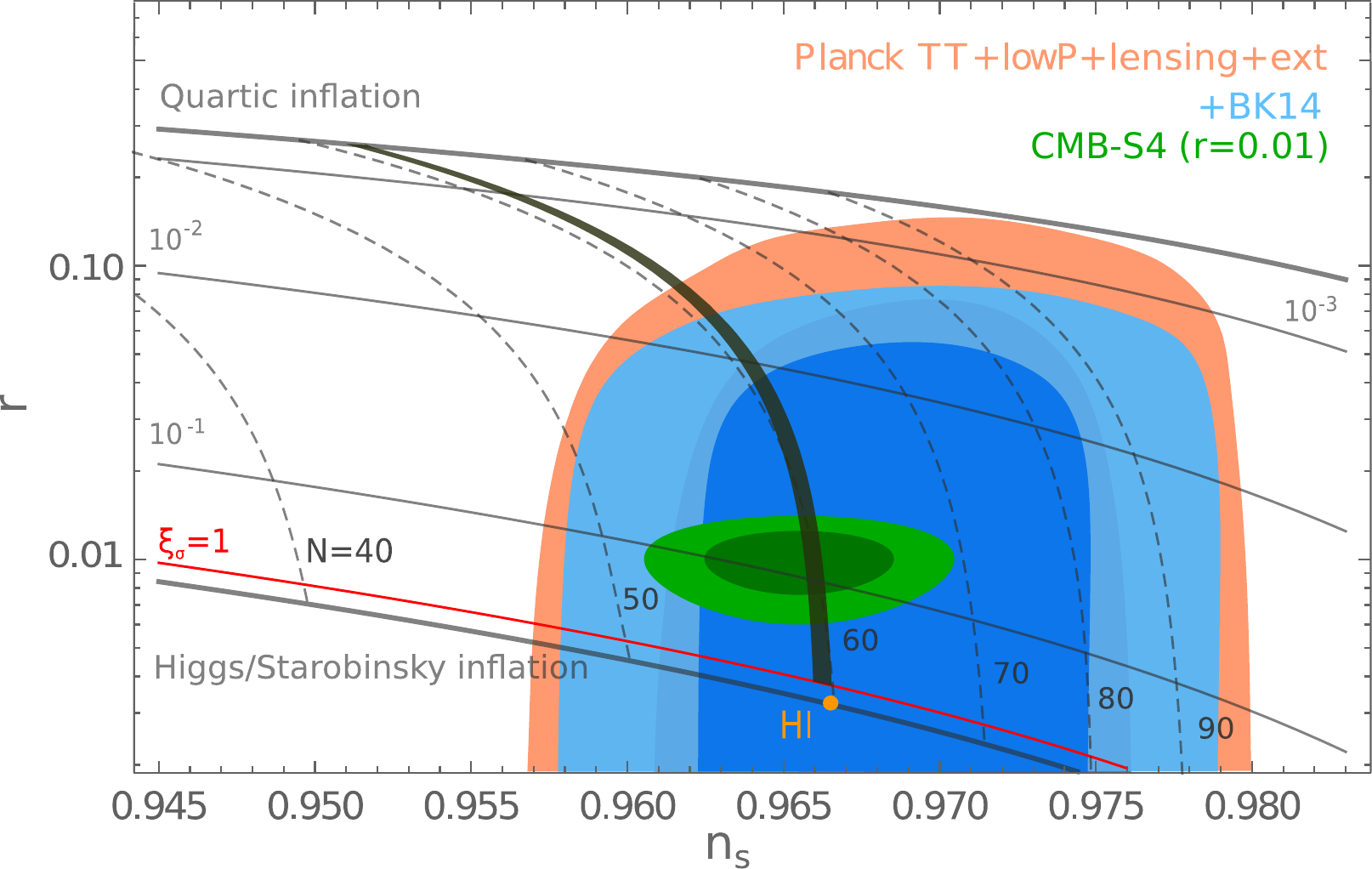}
\caption{\label{fig:r_vs_ns}  \small 
{Predictions for the potential of equation \eqref{genpotential}
in the $r$ vs $n_{s}$ plane with a pivot scale of 0.002 Mpc$^{-1}$. Contours of constant 
$\xi_\sigma$ are shown as black solid lines. The SMASH prediction accounting for a consistent reheating history is given by the thick black line, while the thin dotted lines give isocontours of the number of e-folds that ignore reheating constraints. Also shown are the 68\% and 95\% C.L.  regions at 0.002 Mpc$^{-1}$ of ref. \cite{Array:2015xqh} and the projected sensitivity of CMB-S4 \cite{Abazajian:2016yjj} (in green).
The line labelled as ``Quartic inflation" shows the prediction for a quartic potential (corresponding to the limit $\xi_\sigma \rightarrow 0$), while we also show a black solid line corresponding to the limit $\xi_\sigma\rightarrow \infty$, in which the dynamics is analogous to that in the  Starobinsky \cite{Starobinsky:1980te} and Higgs inflation (HI) \cite{Bezrukov:2007ep} models. 
{The HI result of \cite{Gorbunov:2012ns} is indicated as a point on this line. Adapted from Ref. \cite{Ballesteros:2016xej}.}
}}
\end{center}
\end{figure}

Figure \ref{fig:chi2} from Ref. \cite{Ballesteros:2016xej} shows the agreement of the 
non-minimally coupled potential \eqref{genpotential} with the CMB 
at the pivot scale $0.05$ Mpc$^{-1}$ \cite{Ade:2015lrj,Array:2015xqh}, summarized in the 
 the amplitude of scalar perturbations $A_s$,  the spectral index $n_s$, and the 
tensor-to-scalar ratio $r$,  
\begin{eqnarray}
A_s&=&(2.207\pm 0.103)\times 10^{-9},
\\
n_s&=&0.969 \pm 0.004,
\\
r&<& 0.07 \quad .
\end{eqnarray}
Current constraints from the latest Planck analysis (July 2018) are very similar to the ones quoted above \cite{Akrami:2018odb}. Importantly, the effective quartic coupling $\lambda$ has to be small enough, 
$\lambda \lesssim 10^{-10}$, so that the required non-minimal coupling to fit the amplitude of primordial scalar perturbations is at most 
$\xi_\sigma \lesssim 1$, cf. Fig.\ \ref{fig:chi2} (up right). 
In this region of parameter space, the perturbative consistency of HSI and HHSI is guaranteed and superior to Higgs Inflation, which necessarily {operates} at large $\xi_H$ for the measured value of the top mass, since in this latter case the value of $\lambda_H$ as determined from the measured Higgs mass is sizable \cite{Ballesteros:2016xej}. 
The predictions of the potential \eqref{genpotential}} in the case $\lambda=\lambda_\sigma$ {(or $b\rightarrow 1$ in HHSI)} for { the tensor-to-scalar ratio $r$} vs. the scalar spectral index $n_s$ are shown in FIG. \ref{fig:r_vs_ns} for various values of $\xi_\sigma$. The requirement of predictive inflation, free of unitarity problems, demands $r\gtrsim 0.01$, which will be probed by the next generation of CMB experiments such as CMB-S4 \cite{Abazajian:2016yjj}, LiteBird \cite{Matsumura:2013aja} and the Simons Observatory \cite{Ade:2018sbj}. Since in SMASH and its extensions the particle content is known, the reheating process can be computed in detail. This allows to constrain $n_s$ and $r$ to a narrow band, unlike for generic inflationary potentials devoid of a connection to the SM.

The generalisation of eq.~\eqref{effective_quartic_couplings} to the case of 
a 2HDM --as relevant for the 2hdSMASH model-- or to even more scalars --as relevant for gutSMASH-- has not been worked out yet in full generality. For the related non-minimal Higgs Inflation 
in the 2HDM, see Ref.   \cite{Gong:2012ri}.  However, as far as HSI inflation is concerned, i.e. as long as the non-minimal couplings of all scalars apart from the saxion can be neglected, it is clear that the relevant potential for inflation is --in the Einstein frame--  identical 
 in SMASH HSI. Correspondingly,  in this case, the same inflationary predictions as in SMASH HSI apply also 
for 2hdSMASH and gutSMASH HSI.

\section{STABILITY}
\label{sec:stability}

Primordial inflation of the kind described in the previous section is driven by a positive potential energy and Planckian field excursions. Therefore, a consistent realization within SMASH-type models requires a positive effective potential all the way up to the Planck scale.  Although classical dynamics during inflation only requires a positive effective potential along the inflationary trajectory, instabilities in other regions of field space are dangerous because the fields can end up trapped in them as a result of the quantum fluctuations generated during inflation. To avoid this issue altogether we can demand a strictly positive potential in all field directions. 
Such requirement of (absolute) stability is threatened in the SM by  loop corrections to the Higgs potential due to the top quark. When capturing virtual corrections by means of an RG-improved effective potential with parameters that run with the field scale ($\mu\propto h$), an instability  arises for the preferred values of the Higgs and top masses as a result of  negative contributions to the beta function of the Higgs quartic coupling. In SMASH(d/u) (cf. Table \ref{tab:couplings_ksvz}) --with a portal interaction between the Higgs and the complex scalar $\sigma$ containing the axion-- one can circumvent this problem thanks to the threshold stabilisation mechanism pointed out in Refs.~\cite{Lebedev:2012zw,EliasMiro:2012ay}. In the presence of the Higgs portal coupling, with the $\sigma$ field acquiring a large VEV, the relation between the Higgs mass and the Higgs quartic coupling is altered with respect to that in the SM, such that the quartic can be larger in SMASH than in the SM. At an appropriate matching scale $\mu_0$, the couplings in SMASH and the SM are related as
\begin{align}
\label{eq:threshold}
 \lambda_H(\mu_0)=\lambda_H^{\rm SM}(\mu_0)+\delta(\mu_0), \quad \delta\equiv\frac{\lambda_{H\sigma}^2(\mu_0)}{\lambda_\sigma(\mu_0)}.
\end{align}
Despite the larger value of $\lambda_H$, stabilization is a bit subtle because, as expected from the decoupling of the massive $\sigma$ field at low scales, the SM potential with its corresponding quartic can always be recovered in an appropriate region of field space. For $\lambda_{H\sigma}>0$ this region is of limited extent and can be made not to reach the SM instability scale. Then the potential in the SM-like region can stay positive, while outside of it the larger value of $\lambda_H$ can ensure stability up to Planckian scales. Stabilization is then a tree-level effect and requires a small enough $v_\sigma$ (which is harder to realize in GUT models), in order to ensure that the SM-like region does not go beyond the scale of the SM instability.  For $\lambda_{H\sigma}<0$ on the other hand the SM-like region of the potential extends to arbitrary scales, and stabilization must crucially rely on loop effects that correct the running of the effective quartic coupling in the SM-like region. Stability can be achieved thanks to the positive contributions to the beta function of $\lambda_H$ proportional to $\lambda_H$ itself, which can counter-balance the negative corrections from the top quark: while in the SM the effect of the $\lambda_H$-dependent corrections is sub-dominant, this changes in SMASH due to the larger values of $\lambda_H$ ensured by the modified matching in \eqref{eq:threshold}.

Of course, one also needs to guarantee stability in the $\sigma$ direction, which can again be endangered by fermion loops, this time coming from the RH neutrinos and the exotic quark $Q$. In this case stability can be achieved by demanding sufficiently small Yukawas.

After accounting for the previous effects, we have found that for the SMASH model stability in the Higgs direction can be achieved if the threshold parameter $\delta$ in eq.~\eqref{eq:threshold} is roughly between $10^{-3}$ and $10^{-2}$ (for $\lambda_{H\sigma}>0$) or $10^{-3}$ and $10^{-1}$ (for $\lambda_{H\sigma}>0$), depending on the top mass. On the other hand, stability in the 
$\sigma$ direction restricts the Yukawa couplings of the RH neutrinos and $Q$ to
\begin{align}
 \label{eq:stability_sigma}
 \sum_i Y^4_{ii}+6 y^4\lesssim \frac{16\pi^2\lambda_\sigma}{\log\left(\frac{30 M_P}{\sqrt{2\lambda_\sigma}v_\sigma} \right)},
\end{align}
in the case that the Peccei-Quinn symmetry is extended to a lepton symmetry. Otherwise, the contribution of the Yukawas 
$Y_{ii}$ on the left-hand side of Eq.  \ref{eq:stability_sigma} is absent.

A stability analysis for 2hdSMASH and gutSMASH models  is of course more 
involved due to the extra scalars and has not been done in full generality yet. 

\section{REHEATING}
\label{sec:reheating}

After inflation, the background scalar fields that drove the accelerated expansion will typically oscillate around a minimum of the potential, and throughout these oscillations they will lose energy by producing SM particles that reheat into a  plasma which ends up dominating the energy density of the universe. This reheating process was studied in detail in SMASH \cite{Ballesteros:2016xej}, and arises from the coupled dynamics of the field $\sigma$ containing the axion, the Higgs and the weak gauge bosons. As long as the relevant dynamics only involves Higgses and a complex singlet, and all the other scalar fields remain heavy and decoupled, we expect that some of the features of reheating in SMASH may apply for other variants as well. Differences may arise due to choosing different parameters or from the presence of additional fields with nontrivial dynamics. For example, stability requirements end up enforcing some kinematic blockings in SMASH which could be lifted in other scenarios. And within GUT models, the presence of multiple components within the GUT multiplets containing the axion or Higgses could have nontrivial consequences.

\begin{figure}[t]
\begin{center}
\includegraphics[width=0.7\textwidth]{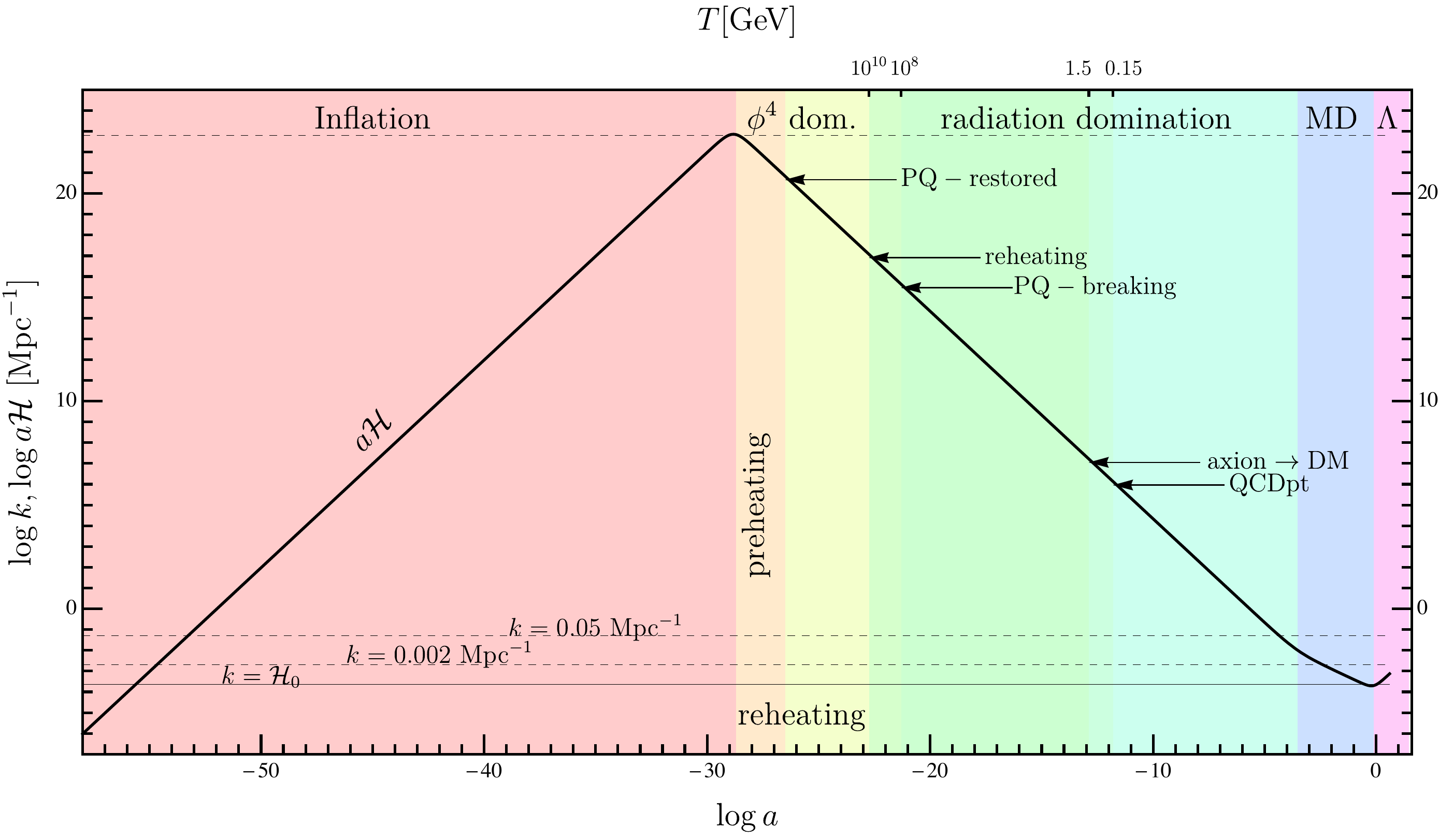}
\caption{\small The history of the Universe in SMASH HHSI, emphasising the transition from inflation to radiation-domination-like Universe expansion $a{\mathcal H}\propto 1/a$ before standard matter and cosmological constant domination epochs \cite{Ballesteros:2016xej}.}  
\label{fig:history}       
\end{center}
\end{figure}

Within the SMASH model, slow-roll inflation ends for $\rho\sim \mathcal{O}(M_P)$, when the inflaton field starts undergoing Hubble-damped oscillations in a quartic potential (for such field values and for $\xi_\sigma\lesssim 1$, as required for predictive inflation, the effect of the non-minimal gravitational coupling can be ignored). These oscillations source a stress-energy tensor whose time-average mimics a radiation fluid. Hence, radiation domination starts right after inflation, and lasts through the phase of reheating in which the oscillating fields trigger the production of SM particles and the energy of the inflaton is transferred into the SM plasma. This post-inflationary history in a radiation-domination era (see  Fig. \ref{fig:history} for a summary of the cosmological history of SMASH) fixes the relation between the scales of the matter perturbations we observe in the Universe today and the size of the primordial fluctuations which gave rise to them, when they outgrew the Hubble horizon and became frozen until their later horizon re-entry.  This relation between scales determines the number of e-folds between a perturbation's horizon crossing and the end of inflation, which fixes the thick black lines
 in Figs. \ref{fig:chi2}, \ref{fig:r_vs_ns} as the prediction for the parameters in SMASH.
 
 In order to understand the process of particle production from the oscillating background field, one has to account for nonperturbative parametric resonance effects \cite{Kofman:1997yn,Tkachev:1998dc}. When the background field changes slowly in time --away from successive crossings of the origin-- one can describe the fields through an adiabatic approximation in which particle number is well defined, and conserved. However, during the crossings the adiabatic approximation breaks down and the appropriately matched adiabatic solutions separated by a crossing have different particle numbers. This particle production is dominated by bosonic fields, and can be understood as a resonance effect accounting for many-body bosonic interactions. The oscillating field may be thought of as a condensate of scalar particles with energy equal to the oscillating frequency, which for a quartic potential goes as 
 \begin{align}
  \omega=\sqrt{\lambda}\,\phi_0,
 \end{align}
 with $\phi_0$ the oscillating amplitude.
 In SMASH, the relevant effective quartic for the inflationary background is determined by $\lambda_\sigma$ --see eq.~\eqref{effective_quartic_couplings}--, which is fixed to $\lambda_\sigma\lesssim 10^{-10}$ by inflationary constraints. In turn, the inflaton condensate couples to Higgs particles with an effective mass dominated by background-dependent contributions, going as $\sqrt{\lambda_{H\sigma}}|\phi|$. Stability constraints on the $\delta$ parameter of eq.~\eqref{eq:threshold} typically imply $\lambda_{H\sigma}\gg\lambda_\sigma$, so that the background Higgs mass is on average much larger than the energy of the particles in the condensate, and Higgs production is blocked except during crossings ($\phi=0$). Due to this, nonperturbative particle production is dominated by the growth of perturbations of the field $\sigma$ itself, for both the real and imaginary part. This effect, confirmed by lattice simulations \cite{Ballesteros:2016xej}, breaks the coherence of the oscillating background and leads to a nonperturbative restoration of the PQ symmetry, as the phase of $\sigma$ ends up taking random values across the Universe. The loss of coherence of $\phi$  ends up further blocking  the production of Higgs particles, as $|\phi|$ stops having an oscillatory behaviour and the Higgs mass always  remains above the frequency of the condensate.
 
In HSI, the Higgs is the only field that couples directly to the inflaton and the production of SM particles is quenched by this effect. The energy of the inflaton gets evenly distributed between the modulus and the phase of $\sigma$, and lattice simulations show that the axion excitations generated in this preheating phase are highly relativistic \cite{Ballesteros:2016xej}. Reheating into SM particles only becomes possible when the $\sigma$ fluctuations redshift below the scale $f_A$, the PQ symmetry becomes broken and the $\rho$ field acquires a mass that finally allows the decay into Higgses. This late decay results in a low reheating temperature of around $T\sim10^7$ GeV, while the initial production of relativistic axions results in an unacceptable amount of dark radiation at late times, predicting an increase in the effective number of relativistic degrees of freedom of
 $\Delta N_\nu^{\rm eff}={\mathcal O}(1)$, which is ruled out by the Planck constraint  $N_\nu^{\rm eff} = 3.04 \pm 0.18$ at 68\% CL 
 \cite{Ade:2015xua}. 
 
In HHSI on the other hand the inflaton is an admixture of $H$ and $\sigma$. This mixing endows the inflaton with a tree-level coupling to gauge bosons. Again, the gauge bosons in the Higgs background acquire oscillating masses $m_W\sim g H\sim g\sqrt{|\lambda_{H\sigma}|/(2\lambda_H)}\phi$ whose average is typically above the frequency of the condensate, but which become zero at the inflaton's crossings of the origin. Crucially, since as argued before the growth of Higgs perturbations is thwarted by the fast production of $\sigma$ excitations, the Higgs component of the background does not lose coherence and continues to oscillate, which keeps the production of electroweak gauge bosons open during crossings. The decay rate of the gauge bosons is fast enough to essentially deplete their population between crossings, so that the boson production is never resonantly enhanced. Nevertheless, a thermal feedback mechanism takes place which enhances the rate of extraction of energy from the inflaton into the SM plasma. The decay products of the gauge bosons quickly reach a thermal bath, which may in turn produce gauge bosons by inverse decays near the crossings. Away from them, the extra bosons gain energy from the condensate as their mass grows with increasing $|\phi|$, and this energy is transferred  into the SM plasma when the massive gauge bosons decay. Modelling this dynamics using Boltzmann equations and energy conservation constraints, one can predict a reheating temperature in HHSI near $10^{10}$ GeV. This implies a thermal restoration of the PQ symmetry, as the critical temperature $T_c$ for the PQ transition goes as
 \begin{align}
 \label{eq:Tc}
  \frac{T_c}{v_\sigma}\simeq\frac{2\sqrt{6\lambda_\sigma}}{\sqrt{8(\lambda_\sigma+\lambda_{H\sigma})+\sum_i Y^2_{ii}+6 y^2}},
 \end{align}
 and $T_c$ is below $10^{10}$ GeV for the preferred SMASH parameters. Moreover, the reheating temperature is also enough to guarantee that the axion population reaches thermal equilibrium, so that its abundance is no longer fixed by the earlier nonperturbative production. In this way the HSI problem with $\Delta N_{\rm eff}$ is avoided, and one predicts a modest amount of cosmic axion backgrond radiation (CAB) corresponding to  $\triangle N_\nu^{\rm eff}\simeq 0.03$, which may be 
probed with future CMB and large scale structure observations \cite{Baumann:2017gkg}.

Within GUT variants, the gutSMASH model with $f_A$ independent of the unification scale could  feature similar dynamics as SMASH in appropriate regions of parameter space. On the other hand, for the miniSO(10)PQ model the large $f_A\gtrsim2.6\times10^{15}$ GeV  can give rise to important differences.\footnote{Similar considerations apply for the miniSU(5)PQ model.}  For example, if the reheating temperature is comparable to that in SMASH, the large value of $f_A$ might mean that a thermal restoration of the PQ symmetry can be avoided, since the critical temperature is proportional to the VEV of the PQ field (see eq.~\eqref{eq:Tc}). This can be a nice feature of the model, as for large $f_A$ one should avoid a post-inflationary restoration of the PQ symmetry in order to avoid overclosure of the Universe by axion dark matter, as reviewed in the next section. However, this still leaves open the possibility of a non-thermal restoration of the PQ symmetry due to the preheating dynamics. Luckily, the large value of $f_A$ can again come to the rescue. The large growth of perturbations in the inflaton field can be hampered for large $f_A$ because the modulus of the field can become quickly trapped around the minimum before the fluctuations in the angular component grow large enough so as to restore the PQ symmetry. Once trapped in the minimum, the $\rho$ fluctuations become massive and can decay quickly into SM particles, so that the growth of angular perturbations is expected to stop.  With the results of the lattice simulations in SMASH  \cite{Ballesteros:2016xej}, one can do a simple extrapolation to estimate the time at which the redshifting oscillations of the field reach a maximum of the order of a given value of  $f_A$. If the time is below the onset of the parametric growth of the angular perturbations, one then expects that PQ restoration will be avoided. Such estimate gives that the PQ restoration might be avoided for $f_A\gtrsim 4\times 10^{16}$ GeV, which is in the allowed window of eq.~\eqref{fgut} and raises the hope that the miniSO(10)PQ model could have a viable parameter space with a consistent cosmological history compatible with pre-inflationary axion dark matter.

\section{DARK MATTER}
\label{sec:dark_matter}

The most important prediction of SMASH  is that the PQ symmetry is broken after inflation. 
In the post-inflationary scenario, dark matter is produced by the re-alignment mechanism~\cite{Preskill:1982cy,Abbott:1982af,Dine:1982ah} \emph{and} the decay of topological defects (axion strings and domain walls) \cite{Kawasaki:2014sqa}. In models where $N_{\rm DW}>1$ 
and the PQ symmetry is exact, there are $N_{\rm DW}$ degenerate CP-preserving vacua and domain walls develop between them when the axion field becomes non-relativistic; i.e.\ when at some temperature $T_1$ the Hubble scale becomes of the order of the axion mass: ${\mathcal H}(T_1)\sim m_A(T_1)$. 
Since there is no preferred vacuum, the system of strings and walls is predicted to continue a scaling regime where the energy in domain-walls soon exceeds the observations.  
Therefore those models have to be discarded~\cite{Sikivie:1982qv} and $N_{\rm DW}$ can only  be 1 in SMASH. This is the main motivation for introducing just one extra heavy quark in SMASH. The alternative models with larger values of $N_{\rm DW}$ can only become viable in scenarios in which the PQ symmetry is not exact --so that the degeneracy of the CP-preserving vacua can be lifted, and the domain-walls become unstable-- or when the PQ symmetry is broken before or during inflation, never to be restored afterwards. In such a situation the energy density stored in the domain walls is simply diluted away by the exponential expansion of the universe during inflation.

Owing to the post-inflationary scenario, the original SMASH model becomes extremely predictive, at least in theory. 
In principle the axion DM abundance in this scenario is calculable by performing numerical simulations of the axion-string-wall network. 
The physics determining axion DM  
depends crucially on $m_A$. Uncertainties from the unknown initial conditions of the axion field are averaged away over many causal domains. 
Since there is no other cold DM candidate in the model, axions should provide all the observed CDM abundance and the theoretical relation $\Omega_Ah^2(m_A) = 0.12$ allows to obtain the required value of $m_A$ (and thus $f_A$). Unfortunately, there is a long-standing controversy regarding the calculation of $\Omega_Ah^2=\Omega_Ah^2(m_A)$. Because of the large dynamical range required 
($f_A/{\mathcal H}(T_1)\sim 10^{30}$ from string cores to the horizon size) an extrapolation is mandatory and different authors have argued differently on how to perform it. Recently, a new method has been developed to endow the strings with the physically motivated effective tension, $\propto \log f_A/{\mathcal H}$, (if not the energy distribution around the string) and has lead to a very precise prediction, $m_A\simeq (26.2\pm 3.4) \mu$eV~\cite{Klaer:2017ond}. The axion DM mass results so small because much of the network energy is radiated in hard axions (which count less for DM) and other hard quanta of the several extra fields that need being introduced. 
A recent detailed study of the string-network evolution \cite{Gorghetto:2018myk} has clarified substantially the results from standard numerical simulations and challenged the results of~\cite{Kawasaki:2014sqa}.  
The authors disregard the effective model of~\cite{Klaer:2017qhr} and highlight the huge uncertainty in the extrapolation to physical string-tensions.

When SMASH was proposed, the most detailed numerical simulations~\cite{Kawasaki:2014sqa} were pointing to $m_A\sim 100\, \mu$eV and the uncertainties where revised to   $50\,\mu\mathrm{eV}\lesssim m_A \lesssim 200 \,\mu\mathrm{eV}$ \cite{Ballesteros:2016xej,Borsanyi:2016ksw}. This corresponded to the range $3\times 10^{10}\,\mathrm{GeV}\lesssim f_A \lesssim   1.2\times 10^{11}\,\mathrm{GeV}$. According to the latest results, the lower limit on $m_A$ could be a factor 2 smaller but the upper limit could be much greater. The next years might be decisive in resolving this controversy as new simulation techniques develope. 
\begin{figure}[t]
\begin{center}
\includegraphics[width=0.8\textwidth]{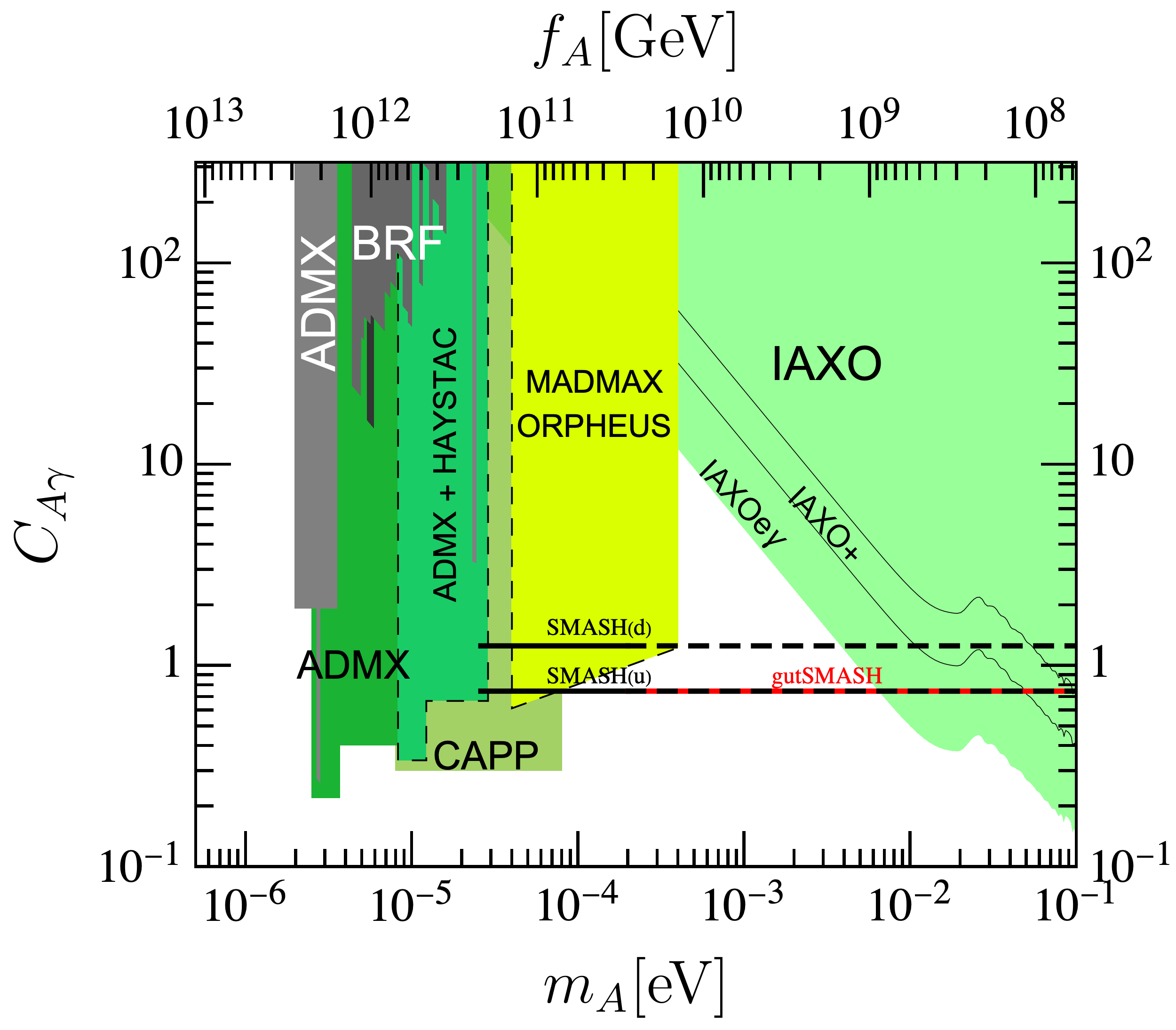}
\caption{\label{fig:pro} \small SMASH predictions for the axion-photon coupling:  SMASH(u,d) (thick solid horizontal lines for hypercharge assignment of $-1/3$ ($2/3$) for $Q$) and gutSMASH (red). The continuing dashed lines show plausible uncertainties. 
We also show, in grey, the current bounds on axion DM (ADMX \cite{Du:2018uak,Boutan:2018uoc},BRF) and, in green, prospects for next generation axion dark matter experiments, such as ADMX~\cite{Du:2018uak,Boutan:2018uoc},  CAPP~\cite{Chung:2018wms}, HAYSTAC~\cite{Zhong:2018rsr}, 
MADMAX~\cite{TheMADMAXWorkingGroup:2016hpc,Brun:2019lyf}, ORPHEUS~\cite{Rybka:2014cya,Morris:1984nu}, and the helioscope IAXO~\cite{Armengaud:2014gea} (fiducial, extended+ sensitivities to the axion-photon channel and IAXOe$\gamma$ for the electron-photon channel). Adapted from Ref. \cite{Ballesteros:2016xej}.}
\end{center}
\end{figure}

Most importantly, this axion dark matter mass window will be probed  in the upcoming decade by axion dark matter direct detection experiments such as ADMX~\cite{Du:2018uak,Boutan:2018uoc}, CAPP~\cite{Chung:2018wms}, HAYSTAC~\cite{Zhong:2018rsr}, RADES~\cite{Melcon:2018dba}, 
 MADMAX~\cite{TheMADMAXWorkingGroup:2016hpc,Brun:2019lyf}, ORPHEUS~\cite{Rybka:2014cya,Morris:1984nu} and others, 
cf. Fig. \ref{fig:pro}. A review on axion DM experiments can be found in~\cite{Irastorza:2018dyq}.

As anticipated earlier, non minimal versions of SMASH where the degeneracy between $N_{\rm DW}$ vacua is broken are in principle possible and can be viable.  Indeed, the degeneracy breaking generates a pressure between vacua that leads to the early collapse of the wall network~\cite{Sikivie:1982qv}. Reference~\cite{Ringwald:2015dsf} studies how fundamental discrete symmetries can be invoked to protect the PQ symmetry from too large a breaking and estimates reasonable phenomenological parameters. This mechanism allows to avoid the domain wall problem for models like an extension of SMASH by further heavy quarks, 2hdSMASH and gutSMASH  within a well motivated framework. The price is however the non-minimality of the extra fields and the discrete symmetry. The best candidates tend to be $Z_N$ symmetries with large $N\sim 9,10$ and point to axion masses in the meV mass ballpark. These predictions do not include the latest results about the string-network evolution that we mentioned above.

If the axion mass is around the meV ballpark, IAXO~\cite{Armengaud:2014gea} could find the concomitant flux of solar axions but direct DM detection will be very difficult. 
The solar signal can be however used to pinpoint the axion mass and couplings~\cite{Dafni:2018tvj,Jaeckel:2018mbn}, thus constraining the SMASH scenario and ease the 
search for DM.  

The post-inflationary scenario typically favoured in SMASH has many interesting phenomenological consequences. A large part of the DM is thought to be in the form of axion miniclusters~\cite{Kolb:1993zz,Kolb:1994fi}, 
small DM halos of typical radius $\sim 10^{12}$ cm and mass $\sim 10^{-12} M_\odot$ that form around matter-radiation equality with large densities $\sim 10^7$ GeV/cm$^3$. 
A recent study shows that smaller and denser objects are also unavoidable and more numerous~\cite{Vaquero:2018tib}. 
Axion miniclusters could be identified with pico-, femto-\cite{Kolb:1995bu,Zurek:2006sy} and micro-lensing~\cite{Fairbairn:2017sil,Fairbairn:2017dmf}, see also~\cite{Katz:2018zrn}. In many cases they will develop solitonic cores, sometimes called dilute axion stars~\cite{Visinelli:2017ooc} when considered in isolation. 
Most axion miniclusters survive until today and are so small that a direct encounter with the Earth is very rare. However, some others are tidally disrupted into streams whose encounters with the Earth can be more frequent and profitable for direct detection~\cite{Dokuchaev:2017psd}. 
The encounters of axion miniclusters/axion stars with the magnetic fields of compact objects has been speculated to be the origin of some fast-radio-bursts~\cite{Tkachev:2014dpa,Iwazaki:2014wka}. 

In general, it is unfortunately impossible to predict whether SMASH variants will always realise the post-inflationary scenario. 
There is a strong tendency for this to be the case also in 2hdSMASH and gutSMASH if all the couplings are small and the inflaton is related to the PQ field. 
The addition of extra fields or non-minimal couplings could 
affect the isocurvature constraints from Planck and the reheating temperature. For the miniSU(5)PQ and miniSO(10)PQ models, as commented in Sect.~\ref{sec:reheating}, the large values of $f_A$ could in principle prevent the restoration of the PQ symmetry --as needed for the extremely light axion to remain compatible with dark matter-- but dedicated studies are needed.

In the pre-inflationary scenario, the PQ symmetry would not be restored and the initial condition of the 
axion field would be an homogeneous local-Universe-wide value that could be anthropically selected for a very broad range of decay constants~\cite{Tegmark:2005dy}. For the axion to furnish all dark matter and $f_A\lesssim3\times10^{17}$ GeV, the initial mis-alignment angle $\theta_I$ has to satisfy \cite{Ballesteros:2016xej}
\begin{equation}
\label{thetaIc}
\theta_{I,c}  \approx 0.0006 \times
 \left(\frac{f_A}{3\times 10^{17}\rm GeV}\right)^{-0.504}.
\end{equation}

We conclude this section discussing DM isocurvature bounds. If the PQ scalar is responsible for inflation one expects that the axion, its angular degree of freedom, gets its quantum fluctuations stretched to superhorizon length scales. Since axions constitute the DM, these fluctuations would get imprinted in the temperature anisotropies of the CMB as an isocurvature component, which is severely constrained by the data \cite{Ade:2015xua}. 
The isocurvature bound gets translated into an upper bound on the Hubble expansion rate ${\mathcal H}_I$ during inflation  (and in turn on the tensor-to-scalar ratio, $r$) as a function of $f_A$. Since there is an upper limit on $r$ from the CMB (see Section \ref{sec:inflation}), this means a maximum possible value of $f_A$. Notice that this bound also depends on the initial axion mis-alignment angle, which together with $f_A$ is the relevant parameter that determines the DM abundance in this scenario of PQ breaking during inflation. 
In scenarios in which the reheating temperature is such that the PQ symmetry becomes restored, the field values of the axion become processed by the thermal (or non-thermal) sub-horizon dynamics and all field perturbations end up being determined by a unique effective temperature scale and of the curvature type; thus, no isocurvature perturbations are generated.

In SMASH and its variants, the energy scale of inflation is mostly determined by the non-minimal coupling $\xi$, which imposes a lower bound on $r$; see figure \ref{fig:chi2}. The PQ symmetry is broken during inflation due to the time-dependent value of $\rho$ --the modulus of the PQ scalar--, which is not at the minimum of its potential, and thus the usual isocurvature bounds do not apply directly, see also \cite{Fairbairn:2014zta}. The reason can be understood by noticing that during inflation the effective $f_A$ ``seen'' by the fluctuations in the direction orthogonal to the inflationary trajectory is actually the instantaneous value of the inflation field. 
Indeed, the ``effective" value of $f_A$ relevant to the isocurvature bounds is larger than the low-energy value of $f_A$ (the one determined by the minimum of the PQ potential, entering into the axion mass equation) thanks to the non-minimal coupling and thus the ensuing constraints get weaker. 
A detailed calculation shows that the maximum allowed value of $f_A$ is $\sim 10^{14}$ GeV \cite{Ballesteros:2016xej}. 
This constraint, together with the fact that the PQ symmetry is always restored for $f_A \lesssim 4\times 10^{16}$ GeV, implies that the only viable SMASH realizations are those 
with PQ restoration after inflation, so that the DM abundance comes not only from oscillations of the axion field but also from the decay of topological defects, as discussed above.

The previous isocurvature bound in principle rules out the viability of miniSU(5)PQ or miniSO(10)PQ, with $f_A$ tied to the unification scale. However, there is a possibility that the bound may be circumvented if one accounts for the fact that the axion field is not really massless during inflation, in contrast to what was assumed when deriving the isocurvature bound described above. During inflation the scalar fields do not sit at their minimum and Goldstone's theorem does not apply; a detailed study of the evolution of the axion mass during and after inflation is needed. 
Moreover, in these models additional fields exist, which opens the possibility for additional paths in field space and further suppression of the bounds. 

\section{BARYOGENESIS}
\label{sec:baryogenesis}

In SMASH models, the presence of right-handed neutrinos with masses proportional to the axion decay constant allows to explain the baryon asymmetry of the Universe through the mechanism of thermal leptogenesis \cite{Fukugita:1986hr}. This relies on out-of-equilibrium, CP-violating decays of heavy RH neutrinos, which generate a net lepton asymmetry which is partly converted into a baryon asymmetry by nonperturbative sphaleron processes that violate baryon plus lepton number. In  SMASH-type models in which the PQ symmetry is restored thermally, such as the HHSI variant of SMASH, the RH neutrinos are massless after reheating, and are expected to acquire thermal equilibrium abundances. After the PQ phase transition they gain a mass, and as long the latter is smaller than the critical temperature of the transition, the massive  RH  neutrinos will typically re-enter equilibrium \cite{Shuve:2017jgj} and decay at later times, generating the asymmetry after inverse decays become Boltzmann suppressed. This scenario is realized in SMASH, where demanding a stabilized potential in the $\sigma$ direction, and assuming  a hierarchy of Yukawas $Y_{22}=Y_{33}=\kappa Y_{11}$ and $y=Y_{11}$, one has
\begin{align}
\frac{T_c}{M_1}\gtrsim \frac{1}{\pi}\sqrt{\left(\frac{2+6\kappa^4}{7+2\kappa^2}\right)\log\left(\frac{30M_P}{\sqrt{2\lambda_\sigma}f_A}\right)},
\end{align}
which follows from eqs.~\eqref{eq:stability_sigma}, \eqref{eq:Tc} and is above 1 for typical SMASH parameters, including the case of near degenerate RH neutrinos with $\kappa\approx1$.

In SMASH realizations in which the PQ symmetry is not restored thermally, as could be the case in models with very large $f_A$, such as GUT variants with $f_A$ correlated with the unification scale, notably miniSU(5)PQ and miniSO(10)PQ,\footnote{Note that in order to avoid problems like monopole production, the reheating temperature in GUTs should be below the unification scale.} the RH neutrinos are massive after reheating, but a thermal initial abundance can still be achieved for a reheating temperature above the RH masses. In this case the asymmetry will again be generated during late-time decays. A thermal initial abundance might not be achieved if the Yukawas of the RH neutrinos are very small, but in these so-called ``weak washout'' scenarios one can still produce an asymmetry from the out-of-equilibrium production and decays of RH neutrinos.

In the vanilla realizations of thermal leptogenesis with hierarchical RH neutrinos, the requirement of a large enough source of CP-violation in RH neutrino decays gives a lower bound $M_1\gtrsim 5\times10^8$ GeV \cite{Casas:2001sr,Giudice:2003jh,Buchmuller:2004nz}. However, since the RH neutrino masses are proportional to their Yukawas with the field $\sigma$, and since these couplings tend to generate destabilizing corrections for the potential in the $\sigma$ direction, having such heavy RH neutrinos can be in conflict with the requirement of stability. For example, in SMASH the stability bound in eq.~\eqref{eq:stability_sigma} for a hierarchical $N_i$ spectrum ($M_3=M_2=3M_1$) requires 
$M_1\lesssim 10^8\, (\lambda/10^{-10})^{1/4} (v_\sigma/10^{11} {\rm GeV})$\,GeV,  which is just borderline 
compatible with the leptogenesis bound.
Nevertheless, leptogenesis can occur for smaller masses with a mild resonant enhancement \cite{Pilaftsis:2003gt} for a less hierarchical RH neutrino spectrum, which relaxes the stability bound and ensures that all the RH neutrinos remain in equilibrium after the phase transition. The estimated level of degeneracy needed in order to reconcile leptogenesis with the stability bound is of the order of 4\%

Finally, even though the RH neutrino masses are typically expected to be heavy, as they are proportional to the axion decay constant, fine-tuned values of the Yukawa couplings still allow for $O(\rm GeV)$ masses. In such cases one recovers the $\nu$MSM at low energies, and even though lepton number violation is suppressed due to the small masses of the RH neutrinos, the baryon asymmetry can arise as a result of out-of-equilibrium oscillations of the right-handed neutrinos \cite{Akhmedov:1998qx}. These give rise to flavoured lepton asymmetries, which may even add up to zero initially, but as long as one flavour is out-of-equilibrium the washout will be incomplete and a net asymmetry will survive.

\section{CONCLUSIONS}
\label{sec:conclusions}

We have provided an overview of SMASHy extensions of the Standard Model which feature a new mass scale 
$v_\sigma$ --of the order of $10^{11}$ GeV in the simplest models, but which could also be tied to a Grand Unification scale around $10^{16}$ GeV-- and provide a falsifiable framework that  addresses the following problems in particle physics and cosmology: inflation, baryogenesis, neutrino masses, dark matter and the strong CP problem. In addition, these models stabilize the electroweak vacuum. Whenever the dynamics of the most economical model  \cite{Ballesteros:2016euj,Ballesteros:2016xej}, called SMASH in this review, is also realized in other extensions (as may happen if the additional fields remain decoupled during inflation and reheating), the models reviewed here
predict a tensor-to-scalar-ratio $r\gtrsim0.004$, a running of the spectral index $\alpha\gtrsim-8\times 10^{-4}$, see figures \ref{fig:chi2} and \ref{fig:r_vs_ns}, and a deviation in the effective number of relativistic neutrino species $\Delta N_{\nu\rm eff}\sim0.03$, values which can be probed in future CMB experiments, such as CMB-S4, LiteBIRD and the Simons Observatory. The SMASH model predicts a lower bound on the axion mass $m_A \gtrsim 25\,\mu{\rm eV}$, in the reach of future axion experiments such as CAPP, MADMAX, ORPHEUS, and IAXO, see figure \ref{fig:pro}. Given that the axion population in the model, constituting the totality of the DM, arises from the re-alignment mechanism and from the decay of topological defects (due to the post-inflationary breaking of the the PQ symmetry), a large fraction of it may be in axion miniclusters, whose abundance may be tested via lensing studies of different astrophysical sources. 

The models surveyed here revolve around the idea of exploiting the complex scalar field that implements the PQ symmetry and solves the strong CP problem. The axion --the angular part of this field-- dynamically relaxes the theta parameter of QCD to a small maximum value, compatible with the upper bounds on the neutron electric dipole moment. On the other hand, the  oscillations of the axion around the minimum of its potential constitute a condensate that can explain the nature of DM. 

The modulus of the PQ scalar is instead the key ingredient for successful inflation. The inflationary sector of SMASH (which also contains a small Higgs component) predicts a primordial spectrum in agreement with the CMB, reheats the Universe efficiently and leads to a small relic abundance of thermal axions which may be identified through a determination of the effective number of relativistic species at early times. The coupling between the Higgs doublet and the PQ scalar is instrumental for the stabilization of the effective potential at large field values, which in the SM is threatened by the large effect on the running of the Higgs quartic coupling coming from the top Yukawa. This interplay between inflation and stability set apart SMASHy extensions of the SM from models which utilize the Higgs alone to drive inflation (an idea that has more severe consistency issues related to the breakdown of perturbative unitarity). 

The small masses of the light neutrinos are explained via the see-saw mechanism, adding three extra right-handed neutrinos whose heavy masses are induced by the VEV, $v_\sigma$, of the PQ scalar, which is proportional to the axion decay constant $f_A$. These heavy neutrinos can also explain the matter/anti-matter asymmetry of the Universe via thermal leptogenesis. The particle content of SMASH is illustrated in figure \ref{fig:SMASH}. In addition to the PQ scalar and the three right handed neutrinos, the model features a heavy vector-like quark $Q$ which is required for the KSVZ-like implementation of the PQ symmetry. At sufficiently low energy the model reduces to the SM augmented by small neutrino masses and the axion, $A$. 

Possible extensions of the minimal SMASH model include implementations in Two-Higgs-Doublet models featuring a DFSZ axion, as well as embeddings of the latter into $SU(5)$ and $SO(10)$ GUTs. As long as one of the Higgses and the extra particles in the GUT multiplets are decoupled during inflation, one can expect to recover the inflationary predictions in SMASH. A similar post-inflationary history may be also recovered for an axion decay scale as in SMASH, i.e. near $10^{11}$ GeV. However, for GUTs with the axion scale $f_A$ tied to the unification scale, as in the miniSU(5)PQ and miniSO(10)PQ models, there can be important differences. First, isocurvature axion perturbations generated during inflation might be incompatible with Planck limits; although Ref.~\cite{Ballesteros:2016xej} discarded $f_A>1.4\times10^{14}$ GeV on this account, the bound neglected the nonzero mass of the axion during inflation (arising from the fact that the scalar field is not at its minimum), and this needs to be accounted for. On the other hand, a large $f_A$ is only compatible with axion dark matter in a scenario in which the PQ symmetry is not restored after inflation. Although dedicated lattice simulations are still lacking,  there are indications that such behaviour is possible, as very large values of $f_A$ change the reheating dynamics and quench the generation of axion perturbations.

Given the lack of compelling new physics signals at the LHC, the idea of attempting to tackle several fundamental physics problems together in a simple (but coordinated) manner is appealing. Perhaps, one of the main take home messages from the SMASHy extensions of the SM that we have reviewed here is that the QCD axion might be a hint not only to dark matter, but also to inflation. In our opinion, it is interesting to continue exploring in the future possible connections between seemingly unrelated problems in particle physics and cosmology. 

There exist other recent proposals which are also inspired by minimality and try to address simultaneously several of the SM standing issues. We will mention some of them briefly in the following. The model of \cite{Salvio:2015cja} has the same particle content as the one proposed in \cite{Dias:2014osa} (and the same as in SMASH). It also attempts to address the same five problems of the SM as SMASH, but it differs from it mainly regarding the heavy neutrino masses (which are not sourced by the VEV of the PQ scalar) and, also inflation, which in this case is driven by the Higgs and thus generically suffers from the unitarity issue. It has been recently argued in \cite{Salvio:2018rv} that the model can be safe from this problem if the top and Higgs masses are tuned in such a way that the quartic Higgs coupling relevant at the energies of inflation is very small. The proposal of \cite{Ema:2016ops} aims to explain --in addition to the issues that SMASHy extensions of the SM deal with-- the flavour structure of masses and mixings in the SM. The model differs from SMASH at several points. For example, the origin of the $SU(3)$ anomaly of the PQ symmetry is unspecified. A key assumption in the model is a pole in the kinetic term of the new scalar field, which leads to an asymptotically flat potential after canonical normalization; see e.g.\ \cite{Galante:2014ifa}. It has been argued that this kind of Lagrangian also suffers from an early breakdown of perturbative unitarity, and thus of consistency \cite{Kehagias:2013mya}. The same idea of using a single $U(1)$ symmetry for the flavour and the strong CP problems  was independently proposed in \cite{Calibbi:2016hwq}, although this paper does not deal with inflation nor with the matter/anti-matter asymmetry. A very different kind of proposal has been recently put forward in \cite{Gupta:2019ueh}. This model aims to solve the same problems as SMASH, except inflation, and in addition it tackles the hierarchy problem. It does so by means of the relaxion mechanism \cite{Graham:2015cka} (for the hierarchy problem) and the Barr-Nelson mechanism \cite{Nelson:1983zb,Barr:1984qx} (for the strong CP problem). Baryogenesis is triggered in this case by oscillations of the relaxion field around its final minimum. 

In summary, we are living in interesting times for particle physics and cosmology, in which simple ideas blended together are providing new theoretical insights and unveiling possible connections between different problems.

\section*{Acknowledgments}

Many thanks to Anne Ernst and Luca Di Luzio for the great collaboration on SMASHy extensions of the SM. The work of GB is funded by a {\it Contrato de Atracci\'on de Talento (Modalidad 1) de la Comunidad de Madrid} (Spain), with number 2017-T1/TIC-5520. It has also been supported by {\it MINECO} (Spain) under contract FPA2016-78022-P and the Spanish MINECO's {\it Centro de Excelencia Severo Ochoa Program} under the grants SEV-2012-0249 and SEV-2016-0597. CB and CT acknowledge support from  the Collaborative Research Centres SFB676 and SFB1258 of the Deutsche Forschungsgemeinschaft (DFG), respectively. The work of AR is partly supported by the DFG under Germany's Excellence Strategy – EXC 2121 ``Quantum Universe" -- 390833306.


\begin{footnotesize}

\end{footnotesize}


\end{document}